\documentclass[runningheads]{llncs}
\usepackage[T1]{fontenc}
\usepackage{graphicx}
\usepackage{listings}
\usepackage{amsfonts}
\usepackage{amsmath}
\usepackage{algorithm}
\usepackage[noend]{algpseudocode}
\usepackage{xspace}
\usepackage{booktabs}
\usepackage{hyperref}
\hypersetup{hidelinks}

\newcommand{\code}[1]{\texttt{#1}}

\newcommand{\rfig}[1]{Fig.\ref{#1}}
\newcommand{\rsec}[1]{Sec.\ref{#1}}

\newcommand{\ralg}[1]{Alg.\ref{#1}}
\newcommand{\fizzer}{\texttt{FIzzer}}
\newcommand{\server}{\texttt{Server}}
\newcommand{\client}{\texttt{Client}}
\newcommand{\instrumenter}{\texttt{Instrumenter}}
\newcommand{\libs}{\texttt{Libraries}}
\newcommand{\target}{\texttt{Target}}
\newcommand{\llvm}{\texttt{LLVM}}
\newcommand{\config}{\texttt{Config}}
\newcommand{\results}{\texttt{Results}}

\newcommand{\pth}[1]{\ensuremath{\overrightarrow{#1}}}
\renewcommand{\orcidID}[1]{{\href{https://orcid.org/#1}{\protect\raisebox{3.25pt}{\protect\includegraphics{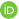}}}}}

\begin{document}
\title{Gray-Box Fuzzing via Gradient Descent and Boolean Expression Coverage}
\subtitle{Technical Report}
%
%
\author{Martin Jon\'{a}\v{s} \inst{1}\orcidID{0000-0003-4703-0795}
  \and
  Jan Strej\v cek \inst{1}\orcidID{0000-0001-5873-403X}
  \and
  Marek Trt\'ik \inst{2,3}\orcidID{0009-0009-6122-9574}
  \and
  Luk\'a\v s Urban \inst{3}\orcidID{0009-0004-9781-3071}
}
%
%
\institute{
  Faculty of Informatics,,
  Masaryk University,
  Czechia\\
  \email{\{martin.jonas,strejcek,trtikm,492717\}@mail.muni.cz}
}
\maketitle              
\begin{abstract}
We present a novel gray-box fuzzing algorithm monitoring executions of
instructions converting numerical values to \code{Boolean} ones. An important
class of such instructions evaluate predicates, e.g., \code{*cmp} in \llvm.
That alone allows us to infer the input dependency (c.f. the taint analysis)
during the fuzzing on-the-fly with reasonable accuracy, which in turn enables an
effective use of the gradient descent on these instructions (to invert the
result of their evaluation). Although the fuzzing attempts to maximize the
coverage of the instructions, there is an interesting correlation with the
standard branch coverage, which we are able to achieve indirectly. The
evaluation on Test-Comp 2023 benchmarks shows that our approach, despite being a
pure gray-box fuzzing, is able to compete with the leading tools in the
competition, which combine fuzzing with other powerful techniques like model
checking, symbolic execution, or abstract interpretation.
\end{abstract}

\section{Architecture}\label{sec:architecture}

\begin{figure}
    \centering
    \includegraphics[width=6cm]{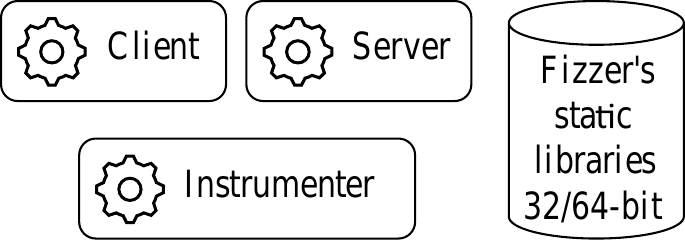}
    \caption{\fizzer's modules}
    \label{pic:fizzer_modules}
\end{figure}

\begin{figure}
    \centering
    \includegraphics[width=\textwidth]{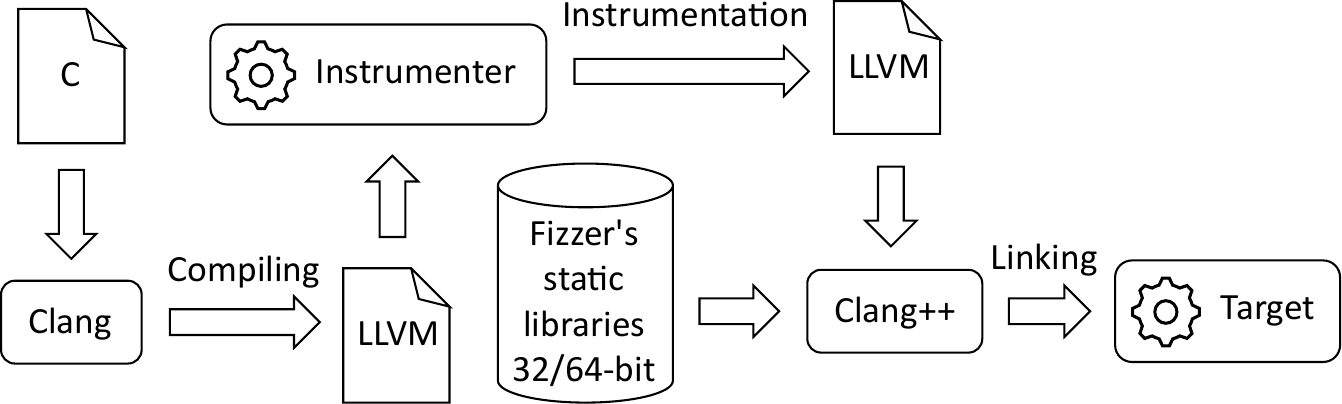}
    \caption{Building the \target{}}
    \label{pic:build_target}
\end{figure}

Our novel gray-box fuzzing algorithm is implemented in a tool called \fizzer{}.
It consists of \server{}, \client{}, and \instrumenter{} 64-bit executables, and
a collection of static \libs{}, each provided in 32 and 64-bit version (see
\rfig{pic:fizzer_modules}%
\footnote{There is also a Python script providing a user friendly interface to
the whole tool.}%
).
The \server{} is responsible for generation of inputs for the analyzed program,
which we denote as the \target{}. It must first be built from an input \code{C}
file into a 32 or 64-bit executable file
\footnote{The \target{} must be build for an architecture with the same endian
as the one used for building of the \server{}.}
, as depicted in
\rfig{pic:build_target}. The \instrumenter{} and the static \libs{} play an
important role in the process. Details are discussed in the next section. 

The \client{} executable mediates communication between \server{} and \target{}
via network. That is an alternative type of the communication. The primary one
is the shared memory. Therefore, \fizzer{} can run without \client{} binary. 
We discuss details of both kinds of communication in \rsec{sec:fuzzing_loop}.

\section{Instrumentation}\label{sec:instrumentation}

The \instrumenter{} is responsible for insertion (instrumentation) of a
monitoring code into the \target{} executable. This code, when executed,
collects valuable data about \target{}'s execution. The data are essential for
an effective input generation in the \server{}.

The \instrumenter{} proceeds in several steps. First, it applies the standard
\llvm{} pass replacing \texttt{switch} instructions by equivalent sequences of
branchings.
\footnote{Next should follow a replacement of calls via pointer by sequences of
branchings, but that is not implemented yet.}
Then, it renames each function in the \llvm{} module such that it adds a prefix
\code{\_\_fizzer\_rename\_prefix\_\_}. This step prevents accidental name
collisions with those in the standard \code{C} library or in \fizzer's \libs{}.

Next, it surrounds each function call instruction 
\footnote{We ignore special functions prefixed by \code{\_\_sbt\_fizzer\_} and
\code{\_\_VERIFIER\_nondet\_}.}
by calls to \fizzer{}'s monitoring functions
\begin{verbatim}
    void __sbt_fizzer_process_call_begin(uint32_t id);
    void __sbt_fizzer_process_call_end(uint32_t id);
\end{verbatim}
both accepting the same unique ID of that call instruction. Tracking function
calls allows \server{} to include calling context into the input generation
process.

Lastly, it inserts monitoring code after each instruction converting one or more
numeric values to a \code{Boolean} one. We call these instructions as
\code{Boolean} instructions. The comparison \code{*cmp} instructions are
\code{Boolean} instructions of the highest importance. However, truncation
instructions and calls to functions returning \code{Boolean} type are also of
the kind. The instrumented monitoring code is supposed to collect maximum
information from the conversion. Namely, conversion is quantified by a value of
the type \code{double}. For truncation and \code{Boolean} function call
instructions the value is always 1. But for a comparison instruction the value
is inferred from its predicate, having a general form $l \bowtie r$, where $l$
and $r$ are some \llvm{} registers of a numeric type and $\bowtie$ is a
comparator from $\{ =, \neq, <, \leq, >, \geq \}$. The instrumented code
computes the value $\code{(double)}l - \code{(double)}r$.
\footnote{If the size of the type of $l$ or $r$ is greater than or equal to the
size of \code{double}, then we may not in fact get maximum information due to
possible overflow or underflow.}
This value is passed as the third argument to the monitoring function
\begin{verbatim}
    void __sbt_fizzer_process_condition(uint32_t id,
            bool instr_result, double value, bool xor);
\end{verbatim}
together with the unique ID of the comparison instruction (1st argument), the
resulting \code{Boolean} value of the comparison instruction (2nd argument), and
\code{Boolean} value determining whether there appears a \code{xor} instruction
anywhere before the comparison instruction in the same basic block or not.

For example, a \code{C} expression
\begin{verbatim}
    x < 123456789
\end{verbatim}
where \code{x} is of the \code{int} type, is expressed by \llvm{}'s
\code{Boolean} instruction
\begin{verbatim}
    %4 = icmp slt i32 %3, 123456789
\end{verbatim}
where \code{\%3} is the register holding the value of \code{x}. The
\instrumenter{} inserts the following code after the instruction 
\begin{verbatim}
    %5 = sext i32 %3 to i64
    %6 = sub i64 %5, 123456789
    %7 = sitofp i64 %6 to double
    call void @__sbt_fizzer_process_condition(i32 1, i1 %4,
              double %7, i1 false)   
\end{verbatim}

At this point it may not be clear, why we instrument \code{Boolean} instructions
rather than branching \code{br} instructions. The reason for that is to be able
to compute the \code{double} values with the maximal precision. For example, if
we instrumented branching instructions, then we would get almost zero precision
from any \code{C} code of this pattern
\begin{verbatim}
    int foo(int b) { ... if (b) ... }
    ... foo(x < 123456789) ...
\end{verbatim}
Observe that \code{foo} accepts \code{int}, which can only be either zero or
one. Therefore, the value \code{(double)b - (double)0}
\footnote{\code{if (b)} is only an abbreviation of \code{if (b != 0)}.}
computed at the branching \code{if (b)} inside \code{foo} can also only be
either zero or one. In contrast, the value \code{(double)x - (double)123456789}
computed from the \code{Boolean} instruction at the call site can be arbitrary.

The same effect can also be observed for another frequently used pattern
\begin{verbatim}
    struct ListItem { ... bool flag; ...  };
    ... item->flag = x < 123456789; ...
    ... if (item->flag) ....
\end{verbatim}
In this code the result of the evaluation of \code{x < 123456789} is stored in a
list item and it is used later in a program branching.

Unfortunately, instrumentation of \code{Boolean} instructions has also a
drawback, related to measuring coverage.
\begin{itemize}
    \item A \code{Boolean} instruction is covered, iff it was evaluated for at
        least one test generated by the \server{} to \code{true} and also for at
        least one test to \code{false}. 
    \item A branching \code{br} instruction is covered, iff it was evaluated for
        at least one test generated by the \server{} such that the execution
        continued to the \code{true} branch and also for at least one test the
        execution continued to the \code{false} branch. 
\end{itemize}
Now, consider the following \code{C} program
\begin{verbatim}
    int x,y;
    ... // Read input to variables x and y.
    bool b1 = (x == 1);
    bool b2 = (y == 1);
    if (b1)
        if (b2) return 1; else return 2;
    else
        if (b2) return 3; else return 4;
\end{verbatim}
If the \server{} generates two inputs
\begin{verbatim}
    x <- 0, y <- 0    and    x <- 1, y <- 1
\end{verbatim}
then both \code{Boolean} instructions \code{x == 1} and \code{y == 1} are
covered, while only branching instruction corresponding to \code{if (b1)} is
covered and the other two are not.

Although \fizzer{}'s primary goal is to generate inputs maximizing coverage of
branching instructions, the goal is approached indirectly through maximizing
coverage of \code{Boolean} instructions. The reasons for that is the fact, that
effectivity of \server{} in input generation fundamentally depends on
information captured in the computed \code{double} values.

The secondary information contributing to the efficiency of input generation is
the count of input bytes read from the start of the \target{} up to each call to
this monitoring function. The count is not passed to the function as the
parameter, because the information is available from functions providing input
to the program (they are discussed below). Therefore, the count of the input
bytes read is recorded together with the information passed via parameters.   

\bigskip

The \libs{} linked to the \target{} provide the main function of the executable
(the original one is renamed and called from the library one), and definitions
of functions called from the instrumented monitoring code, namely:
\begin{verbatim}
    void __sbt_fizzer_process_condition(uint32_t id,
            bool instr_result, double value, bool xor);
    void __sbt_fizzer_process_call_begin(uint32_t id);
    void __sbt_fizzer_process_call_end(uint32_t id);
\end{verbatim}
There are also definitions of functions providing input to the program.
Currently, this is limited to the concept used in the Test-Comp competition,
i.e., to functions with the prototype
\begin{verbatim}
    T __VERIFIER_nondet_T();
\end{verbatim}
where \code{T} stands for any basic type, like \code{int}, \code{char},
\code{float}, etc.

\section{Fuzzing loop}\label{sec:fuzzing_loop}

\begin{figure}
    \centering
    \begin{tabular}{c}
        \includegraphics[width=7cm]{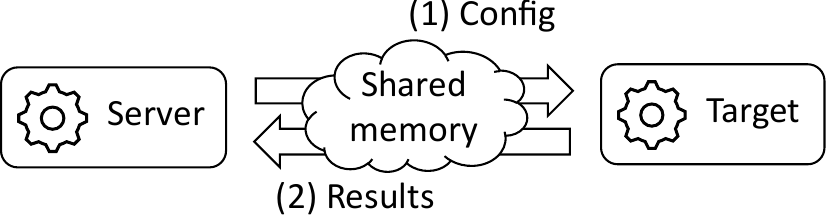}\\\\
        \includegraphics[width=13cm]{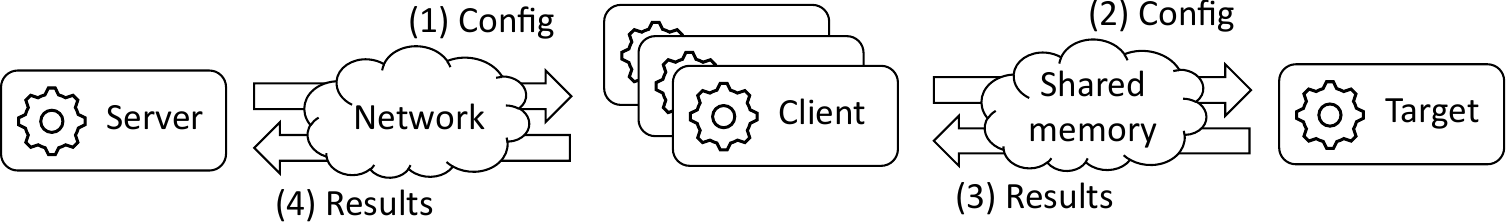}            
    \end{tabular}
    \caption{Fuzzing loop via shared memory (top) and via network (bottom).}
    \label{pic:fuzzing_loop}
\end{figure}

The analysis in \fizzer{} is performed within a top level loop, called fuzzing
loop. In each iteration the \server{} generates an input (which is the subject
of the next section), executes the \target{} with it, and processes data
produced by the executed monitoring code (see previous section).

Details of an iteration, with the focus on data flow, are shown in
\rfig{pic:fuzzing_loop} (top). The \server{} and the \target{} are separate
processes, because the \server{} may generate an input for which the \target{}
crashes. If that crashed \server{} too, the analysis will be over. The processes
exchange data via shared memory, since that is the fastest way of inter-process
communication.

We can further see that \server{} first sends a \config{} to the \target{}.
It comprizes of the following data:
\begin{itemize}
    \item Maximum length of the execution trace. The length is the number of
        executions of the monitoring code of the \code{Boolean} expressions. The
        reason for this limit is simple. Long execution trace consumes a lot of
        memory and its processing by the \server{} decreases an overall
        performance of the analysis.
    \item Maximum stack size. Since our analysis is context sensitive, we also
        restrict size of the stack to manageable size.
    \item Maximum number of input bytes the \target{} may read. The \server{}
        uses inputs from previous iterations of the loop for input generation in
        later iterations. We thus need to keep the size of inputs in reasonable
        bounds so that server can effectively process them.
    \item The name of a model of the input device. There are several types of
        input devices the \target{} program may use, like \code{stdin}, command
        line options, disk, network. \fizzer{} does not work with physical
        devices. A model of a device must always be provided (implemented).
        There can be more models for one device. But currently, there is only
        one model for \code{stdin} device implemented in \fizzer{}. This model
        is initialized with a sequence of bytes, i.e., with the input generated
        by the \server{}. Reading from \code{stdin}
        \footnote{Currently, reading from \code{stdin} can only be done via
        calls to \code{\_\_VERIFIER\_nondet\_} functions (see
        \rsec{sec:instrumentation}).}
        consumes bytes from the sequence. When there is not enough bytes in the
        sequence to be read, then the sequence is automatically extended by
        bytes of a predefined value, which can be either 0 or 85 (there is
        nothing fundamental behind choice of the values). The name of the model
        thus currently primarily determines which of the value should be used. 
    \item A sequence of bytes to be used for the initialization of the model of
        the input device. That is the input generated by the \server{}.
\end{itemize} 

Next, the \target{} reads the \config{} from the shared memory, creates the
model of the device, initializes it with the sequence of input bytes, clears the
shared memory, and calls the original \code{main} function (see
\rsec{sec:instrumentation}). Whenever a monitoring code is executed, it tries to
append the collected data to the shared memory. The execution of the \target{}
always terminates, which happens in these situations:
\begin{itemize}
    \item The \target{} returns from the original \code{main} function. That is
        the normal termination, which the \target{} records in the shared memory
        by setting the termination flag to \code{NORMAL}.
    \item The \target{} executable crashes. This situation is recognized as
        follows. The \target{} sets the termination flag to an invalid value
        before calling the original \code{main} function and to \code{NORMAL}
        once the execution returns from the call. The \server{} always gets the
        exit code from the \target process. If the termination flag is invalid,
        then the \server{} sets the termination flag based on the exit code to
        either \code{CRASH} or \code{NORMAL}
        \footnote{That is for treating forceful termination by calling
        \code{exit(0)} as \code{NORMAL} termination.}
        .
    \item The time reserved by the \server{} for the execution of the \target{}
        was exceeded. In that case the \server{} forcefully terminates (kills)
        the \target{}, and sets the termination flag to \code{TIMEOUT}.
    \item Any of the limits passed to the \target{} in the \config{} was
        exceeded. Then the executions of the \target{} is forcefully terminated
        from within the \target{} by \code{exit(0)} right after setting the
        termination flag in the shared memory to
        \code{BOUNDARY\_CONDITION\_VIOLATION}. We do not distinguish what
        condition was actually violated.
\end{itemize}

The termination flag sits at a reserved location in the shared memory and
represents an important information of the \results{} passed from the \target{}
to the \server{} via the shared memory, see \rfig{pic:fuzzing_loop} (top).
Besides the termination flag the following data are in the \results{} (in the
shared memory):
\begin{itemize}
    \item A sequence of bytes read by the target during the execution. The
        sequence always starts by the input bytes passed from the \server{} to
        the \target{} via the \config{}, but it can be of any length up to the
        limit in the \config{}.
    \item A sequence of types assigned to ranges of bytes in the sequence above.
        A type can be one of the following \code{BOOLEAN}, \code{UINT}$N$,
        \code{SINT}$N$, \code{FLOAT}$M$, \code{UNTYPED}$N$, where $N \in
        \{8,16,32,64\}$ and $M \in \{32,64\}$. For example, if during
        \target{}'s execution there were called functions (in that order):
        \begin{verbatim}
            __VERIFIER_nondet_char();
            __VERIFIER_nondet_float();
            __VERIFIER_nondet_short();
        \end{verbatim}
        then there will be seven bytes in the input bytes sequence. The first
        byte will be associated with the type \code{SINT8}, the range of the
        next four bytes will be associated with \code{FLOAT32}, and the last two
        bytes with \code{SINT16}.
        \begin{remark}
            The types \code{UNTYPED}$N$ are introduced for cases when type
            assignment is not as straightforward as with the use of the functions
            \code{\_\_VERIFIER\_nondet\_}, i.e., when the assignment becomes
            unknown.
        \end{remark}
    \item A sequence of records capturing information about evaluation of all
        \code{Boolean} instructions along the executed path in the \target{}.
        The order of records matches the order of the corresponding
        \code{Boolean} instructions executed along the path. We denote the
        sequence as the execution trace. Each record in the trace consist of the
        following information:
        \begin{itemize}
            \item The unique ID of the \code{Boolean} instruction.
            \item A hash of the calling context. We use context sensitivity in
                order to reduce the number of cases where the \server{} wrongly
                concludes that all reachable \code{Boolean} instructions were
                already considered in the analysis. For example, let us suppose
                we ignore the calling context and we analyze the following
                program:
\begin{verbatim}
    void foo(int x) { if (x < 0) abort(); }
    bool x,y;
    ... // Read input to variables x and y.
    foo(x);
    foo(y);
    ... // A lot of code is here.
\end{verbatim}
                If the \server{} generated, for instance, an input \code{x <- 1,
                y <- -1}, then the \code{Boolean} instruction in \code{foo} will
                be covered. Since no other instruction was discovered, the
                \server{} concludes there is no other reachable \code{Boolean}
                instruction in the program to cover. In contrast, the context
                sensitivity allows us to distinguish the \code{Boolean}
                instruction in each of the two calls of \code{foo}, leaving the
                \code{Boolean} instruction in the second call uncovered. 

                \medskip

                We in fact do not need to know exactly what functions are on the
                call stack. We only want to distinguish \code{Boolean}
                instructions by the contexts. So, we just compute a 32-bit hash
                from IDs (see call site instrumentation in
                \rsec{sec:instrumentation}) of functions on the stack.
                \footnote{Due to recursive functions we restricted computation
                of calling context hash only up to a predefined call stack
                size. For larger context the hash thus remains the same.}

                \medskip

                We denote the unique ID of the Boolean instruction with the
                context hash as an execution ID. 
            \item The result of the evaluation of the \code{Boolean}
                instruction, denoted as direction.
                \footnote{We will see in \rsec{sec:input_generation} that we
                    construct nodes of a binary tree from trace records and the
                    direction identifies the \code{true} or \code{false}
                    successor node in the tree corresponding to the successor
                    record in the trace. I.e., it is the ``direction'' to the
                    successor.}
            \item The \code{double} value, denoted as a value of the branching
                function, computed by the monitoring code from the syntactical
                structure of the \code{Boolean} instruction. For example, for
                \code{*cmp} instructions the branching function is
                $\code{(double)}l - \code{(double)}r$, where $l$ and $r$ are
                registers of a numeric type appearing as arguments in a
                predicate (for details see \rsec{sec:instrumentation}).
            \item \code{Boolean} value determining whether there appears a
                \code{xor} instruction anywhere before the comparison
                instruction in the same basic block or not.
            \item The count of input bytes read from the start of the trace
                (before the first record) up to this record.
        \end{itemize}        
\end{itemize}

The elements of the sequences forming the \results{} are in fact interleaved in
the shared memory. They appear there in the order as the monitoring code in the
\target{} wrote them to the shared memory. Individual sequences are thus
constructed in the \server{} during a sequential scan of the elements the shared
memory.

\bigskip

\fizzer{} implements an alternative version of the fuzzing loop which is
depicted at \rfig{pic:fuzzing_loop} (bottom). We see that, in contrast to the
original version of the fuzzing loop, the \server{} is replaced by a \client{}
binary. The \client{} indeed implements exactly the same procedure of
communication with the \target{}. The \target{} is thus unable to tell whether
it communicates with the \server{} or the \client{}. From the \server{}'s point
of view, the \client{} behaves like the \target{}. Only the communication medium
is different. In summary, the alternative version of the fuzzing loop is a
slower implementation of the original version, because the data flow through two
media, namely the network and shared memory.

The alternative version however can be used in a setup which can potentially
improve the overall performance. Observe in the \rfig{pic:fuzzing_loop} (bottom)
that the \server{} can simultaneously instruct multiple \client s on multiple
computers
\footnote{The alternative version of the fuzzing loop is currently only in a
prototype stage where the loop works only on a \code{localhost} with one
\client{}.}
to execute their \target s. Although the simultaneous executions could be
implemented also in the original version of the fuzzing loop, its practical
applicability is considerably reduced due to limited resources of a single
computer.

\section{Input generation}\label{sec:input_generation}

The goal of input generation is to produce a shortest sequence of inputs for the
\target{} whose executions cumulatively covers the maximum of \code{Boolean}
instructions in the \target{}.

The \server{} initially generates the empty input. All other inputs are
generated by exactly four input generation analyses:
\begin{itemize}
    \item Sensitivity: identifies a subset of input bits, called sensitive
        bits, to be focused on by other analyses.
    \item Bitshare: reuse of sensitive bits in previously generated inputs in
        the construction of new inputs.
    \item Typed minimization: a gradient descent on sensitive bits forming
        variables of a known numeric type.
    \item Minimization: a gradient descent on sensitive bits forming variables
        whose numeric type is not known.
\end{itemize}
They are described in details later in \rsec{sec:input_generators}. Exactly one
of them is active at a time. Only the active analysis generates inputs. Other
analyses wait for their activation. Once an analysis is activated, it stays
active until it either deactivates itself or it is forcefully deactivated. An
analysis deactivates itself, when its input generation strategy is finished. An
analysis is deactivated forcefully, when its goal was achieved before the input
generation strategy is finished.

The goal of all analyses, except the sensitivity, always is to invert the
evaluation result of a particular \code{Boolean} instruction corresponding to a
certain record in an execution trace. The sensitivity analysis has a different
goal - to compute sensitive bits. These bits are essential for all other
analyses. Therefore, we always want to complete its input generation strategy,
i.e., the sensitivity analysis is never forcefully deactivated.

Once an analysis is (forcefully) deactivated, another one must be activated.
That is a responsibility of an analysis selection strategy. The goal of this
strategy is to maximize coverage of \code{Boolean} instructions. It approaches
the problem such that it builds a short-term goals for the four input generator
analyses and activates the analyses for these goals. The ultimate long-term goal
with the maximal coverage is thus achieved indirectly - it is approached by
solving a sequence of short term goals. In the heart of the building short-term
goals there is a maintenance of and a search in core data structures constructed
from the data accepted by the \server{} from the \target{} after each its
execution. We discuss the details of the selection strategy later in
\rsec{sec:generator_selection}. 

In each iteration of the fuzzing loop (see \rsec{sec:fuzzing_loop}) the active
analysis generates exactly one input for the \target{}. The \server{} then
accepts back an input $x$ (which is the generated input, possibly extended or
truncated), the sequence of types $t$ (logically splitting $x$ into sequences of
bits and assigning them types), and an execution trace $T$. These data are used
for construction of core data structures essential for all analyses. It is thus
first necessary to understand these core data structures and how they are built
from the accepted data. That is the subject of the following subsections.

\paragraph{Notation:} If $S$ is a sequence, then $|S|$ denotes the number of
elements in the sequence and $S[i]$ denotes the $i$-th element. We also use
Python-like syntax for denoting subsequences, e.g., $S[k:l]$, $S[:l]$, $S[k:]$,
denote sequences of elements from $S$ at indices $k, \ldots, l-1$, $0, \ldots,
l-1$, $k, \ldots, |S|-1$, respectively. If the element has some structure, then
we use ``dot'' notation to access the fields. For instance, if $T$ is an
execution trace and $0 \leq i < |T|$ is an index to $T$, then following are
all field of yjr record $T[i]$:
\begin{itemize}
    \item $T[i].id$ is the execution ID of the \code{Boolean} instruction
        corresponding to $T[i]$, 
    \item $T[i].f$ is the \code{double} value of the branching function,
    \item $T[i].direction$ is the result of the evaluation of the \code{Boolean}
        instruction,
    \item $T[i].xor$ indicates whether a \code{xor} instruction appears before
        the \code{Boolean} instruction in the same basic block.
    \item $T[i].nbytes$ is the number of input bytes read from the begin of the
        trace (before $T[0]$) up to $T[i]$,
\end{itemize}
Further, fields can be nested, for which we also use the same notation, e.g.,
$T[i].id.uid$ and $T[i].id.ctx$ are the unique ID of a \code{Boolean}
instruction and the context hash, respectively (see \rsec{sec:fuzzing_loop}).
Finally, if we speak about a field of some record in general, we omit the
record, e.g., we just write $id.uid$ when we speak about the unique ID of a
\code{Boolean} instruction.

\subsection{Execution tree}\label{sec:execution_tree}

At the heart of the input generation there is a binary rooted tree, called
execution tree. Initially it is empty. Each node $N$ in the tree corresponds to
an execution of a \code{Boolean} instruction along some program path, for which
the \server{} accepted an execution trace. Since a \code{Boolean} instruction
can be evaluated to two values \code{true} or \code{false}, the node may have
two successors, called \code{true}-successor and \code{false}-successor. Since
the nodes are connected via edges, the node may also have two edges, called
\code{true}-edge and \code{false}-edge. The edges carry labels. We discuss their
purpose later.

\paragraph{Notation:} Let $N$ be a node of the execution tree. Then $\pth{N}$ is
a sequence of nodes in the tree from the root node to $N$ (including $N$). The
depth of $N$ in the tree is the count of edges between nodes in $\pth{N}$ and we
denote it as $dN$. Clearly, $dN = |\pth{N}| - 1$, $\pth{N}[0]$ is the root node,
and $\pth{N}[dN]$ is $N$. When $b$ is a \code{Boolean} value, i.e., \code{true}
or \code{false}, then $N.successor[b]$ is the $b$-successor node of $N$ and
$N.label[b]$ is the label of the $b$-edge of $N$. And $N.parent$ is the parent
node of $N$ in the tree. The parent of the root node is \code{null}.\\

In the end of each iteration of the fuzzing loop the tree is updated according
to data accepted by the \server{} (see \rsec{sec:fuzzing_loop}), which is the
termination flag, an input $x$, types $t$, and a trace $T$. During this process
the tree may be extended (new nodes are created) and some existing nodes may be
updated (their fields).

\subsubsection{Updating tree's shape}\label{sec:update_tree_shape}

A trace $T$ accepted by the \server{} is mapped to the nodes of the execution
tree such that $T[0]$ is mapped to the root node $R$, then $T[1]$ is mapped to
$R.successor[T[0].direction]$, and so on. When $T[i]$ is mapped to a node $N$,
$i + 1 < |T|$, and $N.successor[T[i].direction] = \code{null}$, then the missing
successor node is created and inserted to the tree.

Labels of edges in $N.label$ describe the transition to successor nodes,
including the case the successors are missing. Let $b$ is a \code{Boolean}
value. If $N.label[b]$ is
\begin{itemize}
    \item \code{NOT\_VISITED}, then $N.successor[b] = \code{null}$. This
        indicates that there is no trace among all traces accepted by the
        \server{} so far, which has a record $T[i]$ mapped to $N$ such that
        $T[i].direction = b$. 
    \item \code{END\_EXCEPTIONAL}, then $N.successor[b] = \code{null}$. This
        indicates that there was at least one trace $T$ accepted by the
        \server{}, which has the last record $T[dN]$ mapped to $N$,
        $T[dN].direction = b$, and also the termination flag (see
        \rsec{sec:fuzzing_loop}) of the execution was set to \code{CRASH}.
        Further, there is no trace among all traces accepted by the \server{},
        for which $N.label[b]$ would be set to any of the values listed below.
    \item \code{END\_NORMAL}, then $N.successor[b] = \code{null}$. The
        indication is the same as for the previous label, except the termination
        flag has the value \code{NORMAL}.
    \item \code{VISITED}, then $N.successor[b]$ points to a valid node. That
        indicates there was at least one trace $T$ accepted by the \server{}
        such that $|T| > dN + 1$, $T[dN]$ is mapped to $N$, and $T[dN].direction
        = b$.
\end{itemize}
The values of labels are ordered from top down, i.e., $\code{NOT\_VISITED} <
\cdots < \code{VISITED}$. It favours longer execution paths and also normal
paths over crushes. That maximizes the potential to cover \code{Boolean}
instructions deeper in the code.
A tree node is created with \code{NOT\_VISITED} for both $label$s. The labels
may change during the analysis, namely to increase in that order. For example,
if $N.label[b] = \code{END\_EXCEPTIONAL}$ and the \server{} accepts a trace $T$
such that $|T| > dN + 1$, $T[dN]$ is mapped to $N$, and $T[dN].direction = b$,
then $N.label[b]$ will be changed to \code{VISITED} and $N.successor[b]$ will
point to a newly created node.

\subsubsection{Purpose of branching functions}\label{sec:purpose_of_branching_function}


Let us consider the following C program
\begin{verbatim}
    char x = __VERIFIER_nondet_char();
    ... // some code
    bool bi = x > 254;
\end{verbatim}
The branching function of the \code{Boolean} instruction corresponding to the
variable \code{bi} is $f(x) = \code{(double)}x - 254$. If we want to cover the
\code{Boolean} instruction, then we should attempt to find some inputs $u$ and
$v$ such that $f(u)$ and $f(v)$ have opposite signs. We should first realize the
following:
\begin{itemize}
    \item $f$ may not be linear, because $x$ may not be an independent variable.
        Indeed, the code abbreviated by ``\code{...}'' could modify \code{x}
        arbitrarily. It means $f$ is in fact unknown to us. So, the best thing
        we can do is to sample the function by generating inputs $x$ and
        observing $f(x)$.
    \item Random sampling of the input domain may easily be ineffective for
        obtaining the inputs $u$ and $v$. That can be seen in our example even
        if the code in ``\code{...}'' does not affect \code{x}. Indeed, there is
        only one input evaluating $f$ to a positive number.
    \item Since we search for $u$ and $v$ producing $f(u)$ and $f(v)$ of
        opposite signs, random sampling of the input domain in a neighborhood
        around the global minimum of the function $|f(x)|$ may actually be
        effective. That can be seen in our example, when the code in
        ``\code{...}'' does not affect \code{x}. If we randomly sample the
        inputs from a small neighborhood around the global minimum $254$, then
        our chances of generating the desired inputs $u$ and $v$ quickly will be
        considerably higher (depending on the size of the neighborhood we sample
        from).
\end{itemize}
The purpose of a branching function $f(x)$ is thus to allow us quickly converge
to a neighborhood around the global minimum of the function $|f(x)|$, where we
can then effectively obtain the desired inputs via random sampling form the
neighborhood. We use the gradient descent as the convergence method, where we
compute partial derivatives numerically, since the function $f$ is unknown.
Instead of detecting whether we already are in a neighborhood for an effective
random sampling or not we rather take multiple samples in each gradient descent
step. This way we also take several samples from the neighborhood in the end of
the descent, in a price of taking samples outside the neighborhood.
We discuss details later in \rsec{sec:typed_minimization}.

\subsubsection{Updating content of nodes}\label{sec:updating_nodes}


Let us consider a node $N$. During the analysis the \server{} may accept several
inputs $x_1, x_2, \ldots, x_n$, types $t_1, t_2, \ldots, t_n$,  and traces $T_1,
T_2, \ldots, T_n$, where the records at the index $dN$ are all mapped to $N$.
The values $f$ in these records may be different. Which of the triples $(x_j,
t_j, T_j)$ we need for an effective coverage of $N.id$? Since we want to
approach a neighborhood around the global minimum of $|f(x)|$, only one
tuple seems to be sufficient - the one with the smallest $|T_j[dN].f|$. However,
it is quite common that same bit(s) in the input $x_j$ affect values $f$ in
multiple records in $T_j$. We must therefore consider all predecessors of $N$.
So, we use a triple with the smallest value
$$w_N(T_j) = \sum_{i = 0}^{dN} T_j[i].f^2$$
The squares of values in the sum increase (emphasize) the impact of larger
values and they also handle negative values.

\paragraph{Notation:} Each node $N$ has also fields $N.x$, $N.t$, $N.T$ used for
storing values $x_j$, $t_j$, $T_j$, which give the smallest value of $w_N$. We
further abbreviate accesses to fields of $N.T[dN]$ such that we omit
``$.T[dN]$'', e.g., instead of $N.T[dN].f$ we write just $N.f$. And finally, we
say that an execution trace $T$ is mapped to $N$ (or $\pth{N}$), if $|T| > dN$
and for each index $0 \leq i \leq dN$ and $0 \leq j < dN$ we have $T[i].id =
N.T[i].id$ and $T[i].direction = N.T[i].direction$ (or $T[i].id =
\pth{N}[i].id$, $T[0]$ is mapped to the root node, $\pth{N}[j+1] =
\pth{N}[j].successor[T[j].direction]$).\\

Since the \server{} generates the inputs sequentially (we do not have them all
at once), the field $x$, $t$, $T$ may be changed during the analysis. Namely, if
a new triple $(x_{n+1}, t_{n+1}, T_{n+1})$ is accepted by the \server{} such
that $T_{n+1}[dN]$ is mapped to the node $N$, and $w_N(T_{n+1}) < w_N(N.T)$,
then we set all fields $N.x$, $N.t$, $N.T$ to values $x_{n+1}$, $t_{n+1}$,
$T_{n+1}$, respectively.

There are more information stored in each node. However, these fields are
related to individual input generation analyses and the analysis selection
strategy. So, we introduce these fields later.

\section{Input generation analyzes}\label{sec:input_generators}

We already know there are exactly four analyses responsible for input generation
(sensitivity, bitshare, and two minimization analyzes); exactly one of them is
active at time; an analysis may stay active over several iterations of the
fuzzing loop; the active analysis generates a single input in each iteration of
the fuzzing loop and also processes the corresponding trace in the same
iteration.

The sensitivity analysis differs from other three in the sense that its goal for
any node $N$ in the execution tree is to identify a subsets of bits in the input
$N.x$, called sensitive bits. Other analyses then focus only on the sensitive
bits, which considerably improves the performance of these analyses. In other
words, the goal of sensitivity analysis is to boost effectivity of other
analyses rather than aiming to improving the coverage of \code{Boolean}
instructions. That is also the reason why we always start sensitivity analysis
on $N$ before any other analysis.

The goal of all other analyses is to find a missing successor node of a given
node in the execution tree. More precisely, given a node $N$ in the execution
tree and a \code{Boolean} value $b$ such that
\begin{itemize}
    \item the set of sensitive bits of $N$ detected by the sensitivity analysis
        is not empty,
    \item $N.label[b] = \code{NOT\_VISITED}$, i.e., the $b$-successor of $N$ is
        not in the tree yet,
\end{itemize}
the goal of all other analyses is to find an input $x$ so that the obtained
trace $T$ is mapped to $\pth{N}$, $|T| > dN + 1$ and $T[dN].direction = b$.

Observe that neither these three analyses aiming to improving the coverage of
\code{Boolean} instructions. Indeed, the analysis can be asked to find
$b$-successor of a node $N$, whose corresponding \code{Boolean} instruction with
ID $N.id$ was already covered. The only analysis aiming at the coverage of
\code{Boolean} instructions is the analysis selections strategy, whose task is
to choose a node $N$ in the tree and start one of our four analyses on it,
whenever the previously active analysis becomes inactive. We discuss details of
the selection strategy later in \rsec{sec:generator_selection}.

\paragraph{Notation} Observe that each analysis is activated with a certain node
$N$ in the execution tree. We will see later (namely in
\rsec{sec:generator_selection}) that we need to track the information what
analysis was already applied to what node and when. So, we introduce to each
node $N$ \code{Boolean} fields (flags) $N.sa$, $N.ba$, and $N.ma$ indicating
whether the sensitivity analysis, bitshare analysis, and minimization analysis
respectively were already applied to the node or not. Also notice that we do not
distinguish between the two minimization analyses. That is because at most one
of them can be run on a given node. In order to keep track of when the analyses
were applied we introduce integer fields $N.sn$, $N.bn$, and $N.mn$ which we set
to the number of the fuzzing loop iteration. So, whenever an analysis $y \in
\{s,b,m\}$ is (forcefully) deactivated, then the filed $ya$ is set to
\code{true} and $yn$ is set to the current fuzzing loop iteration number. Notice
that we record the last iteration number, in which the analysis was active
(which is typically after tens or hundreds of subsequent iterations). In
general, beside the node $N$, the fields are set in all nodes in the tree which
were changed by the analysis since its activation. The sensitivity analysis
often computes (updates) sensitive bits of several nodes in the tree along the
path from the root node to $N$. So, fields of all these nodes are thus set. All
other analyses modify only the node $N$, so only fields of $N$ are updated.

\paragraph{Notation} We further use the field $N.fn$ to store the number of the
fuzzing loop iteration, when the field $N.f(x)$ was set for the last time.

\paragraph{Fuzzing loop integration:} In this paper we present the analyses from
the algorithmic point of view. In our implementation the algorithms have a
different structure. The actual computation is of course the same. The reason
for the difference is the integration of the algorithms to the fuzzing loop (see
\rsec{sec:fuzzing_loop}). In each iteration of the fuzzing loop two method of
the analysis are called:
\begin{itemize}
    \item \code{generate\_input}: The analysis is supposed to return an input
        for which the \target{} will be executed.
    \item \code{process\_results}: The analysis is supposed to process the
        obtained execution trace $T$.
\end{itemize}
The algorithms thus contain auxiliary variables providing a bookkeeping of of
its the current state so that they can proceed further within calls to the two
functions above.

\paragraph{Fast execution cache:} Input generation algorithms of some analyses
discussed below may occasionally generate an input already generated before.
Rather than complicating the implementation we introduced a cache to these
analyses. The cache work as a map from 64-bit hashes of all generated inputs to
the \code{double} values of the considered branching function. Any generated
inputs is first looked up in the cache and it is executed by the \target{} only
on cache miss. 

\subsection{Sensitivity analysis}\label{sec:sensitivity}

The purpose of this analysis is to boost effectivity of other three analyses.
Namely, given a tree node $N$, its goal is to compute a set of indices of those
bits in $N.x$ having an impact on $N.f$. We call these bits as sensitive bits.
The other analyses may thus focus only on the sensitive bits, i.e., safely
ignore all others.

Since the formal definition of sensitive bits is not intuitive, we start with
an example. Let us consider this C program
\begin{verbatim}
    char c = __VERIFIER_nondet_char();  // read 8 bits
    c = c & 7;  // Set bits at indices 0,1,2,3,4 to 0.
    bool bi0 = ((c ^ 7) * (c ^ 1)) != 0; // Boolean instruction; ID=0
    if (bi0) return; // Return if c is neither 7 nor 1.
    bool bi1 = c > 2; // Boolean instruction; ID=0
\end{verbatim}
From the second line we can immediately conclude that input bits at indices
0,1,2,3, and 4 may not be sensitive (no impact on branching functions), because
they are cleared after read. There are two \code{Boolean} instructions in the
program; they correspond to the variables \code{bi0} and \code{bi1}. They both
operate on inputs, all with the size $m = 8$ bits.

Let us decide whether the input bit at the index $s = 7$ is sensitive for the
first \code{Boolean} instruction or not. For $m = 8$ we have exactly 256
possible inputs $x_0=0, ..., x_{255}=255$ for which the execution reaches and
evaluates the \code{Boolean} instruction. The evaluation is captured in the
record at index $d = 0$ in all execution traces $T_0, ..., T_{255}$
corresponding to the inputs. We can split all pairs $(x_i, T_i)$ into a disjoint
sets $X_f$ according to the equality of the values $T_i[d].f$, i.e., two pairs
$(x_i, T_i)$ and $(x_j, T_j)$ are in the same set, iff $T_i[d].f = T_j[d].f$.
Since the branching function can evaluate only to four values 0, 7, 8, and 15,
there will be four corresponding sets of the pairs. Intuitively, a bit at the
index $s$ should be sensitive, if there exist pairs $(x_i, T_i)$ and $(x_j,
T_j)$ from different sets such that $x_i[s] \neq s_j[s]$. For instance, inputs
$(x_0, T_0) \in X_7$ and $(x_1, T_1) \in X_0$ and $x_0[s] = 0 \neq 1 = x_1[s]$.
So, the bit at the index $s$ should be sensitive. Although this is the result we
want, the condition we formulated is too weak, because the bit at the index 4
would be sensitive too (c.f., $(x_0, T_0) \in X_7$ and $(x_9, T_9) \in X_0$ and
$x_0[4] = 0 \neq 1 = x_9[4]$). Therefore, we must restrict our search to the
``closest'' inputs from different sets. For that can use the Hamming distance:

\medskip

Let us consider two inputs $u$ and $v$ such that $|u| = |v|$. The Hamming
distance $H(u,v)$ is the number of all indices $0 \leq i < |u|$ where $u[i] \neq
v[i]$.

\medskip

\noindent
Observe that $H(x_0, x_1) = 1$ while $H(x_0, x_9) = 2$. Using both, the
intuitive condition and the Hamming distance, we can further decide that bits at
indices 5 and 6 are also sensitive (c.f., $(x_4, T_4),(x_2, T_2) \in X_{15}$ and
$H(x_0, x_4) = H(x_0, x_2) = 1$) while all other bits are not sensitive.

Let us now focus on the second \code{Boolean} instruction (corresponding to
\code{bi1}). This instruction executed only for 64 of all 256 inputs above. In
all of the corresponding traces the instruction corresponds to records at the
index $d = 1$. For 32 inputs $x_1, x_9, x_{17}, \ldots, x_{249}$ the instruction
is evaluated to \code{false} and for all $i$ we have $T_i[d].f = -1$. And for 32
inputs $x_7, x_{15}, x_{23}, \ldots, x_{255}$ the instruction is evaluated to
\code{true} and for all $i$ we have $T_i[d].f = 5$. So, we have two sets
$X_{-1}$ and $X_5$. Observe, that for any $(x_i, T_i) \in X_{-1}$ and $(x_j,
T_j) \in X_{5}$ we have $H(x_i, x_j) \geq 2$. Also, only bits at indices 5 and 6
satisfy both conditions, i.e., they are sensitive (c.f., $(x_1, T_1) \in X_{-1}$
and $(x_7, T_7) \in X_5$ and $x_1[5] \neq x_7[5]$ and $x_1[6] \neq x_7[6]$)).
Observe, the minimal Hamming distance between sets defines also the minimal
number of bits considered as sensitive simultaneously. We are ready to define
sensitive bits formally.

\medskip

Let $0 < m$ and $0 \leq d$ be integers, $x_1, \ldots, x_n$ be finite sequences
of all possible inputs such that $|x_i| = m$ and $T_1, \ldots, T_n$ be finite
sequences of the corresponding traces such that for all integers $0 < i,j < m$,
$0 \leq k \leq d$, and $0 \leq l < d$ we have $|T_i| \geq d$, $T_i[k].id =
T_j[k].id$, and $T_i[l].direction = T_j[l].direction$. In other words, for each
input $x_i$ the \target{} executes exactly the same sequence of $d+1$
\code{Boolean} instructions (we ignore the suffixes of the traces $T_i[d+1:]$).
The bit at an index $0 \leq s < m$ is sensitive at the trace index $d$, iff  
there exist two pairs $(x_i, T_i) \in X_{f_i}$ and $(x_j, T_j) \in X_{f_j}$ such
that $f_i \neq f_j$, $x_i[s] \neq s_j[s]$, and $H(x_i, x_j)$ is equal to the
minimal Hamming distance between $X_{f_i}$ and $X_{f_j}$.

\medskip

Precise computation of sensitive bits can be expensive in practice. The number
of possible inputs to generate grows exponentially with $m$. We, of course,
consider only inputs for which the execution proceeds along the same program
path up to the record at the index $d$ in the traces. However, enumeration of
only such inputs is a hard problem. Further, we do not know the minimal Hamming
distance between the sets $X_f$ in advance (it may decrease with any input we
try). The goal of the sensitivity analysis is thus to compute only an
approximation of the sensitive bits. The set of detected bits may thus contain
some non-sensitive bits (causing a decrease of effectivity of other analyses)
and/or some truly sensitive bit may be missing the set (causing possible
decrease in the overall coverage of \code{Boolean} instructions, because other
analyses may be then unable to invert their evaluation).

\medskip

The sensitivity analysis computes the approximation of sensitive bits as
follows. Let us consider a node $N$ in the execution tree. So, we have $m = 8
\cdot N.nbytes$ and $d = dN$. We also have one pair $(N.x[:m], N.T[:d+1]) \in
X_{N.f}$. Instead of considering all possible pairs from all possible sets $X_f$
we fix the first pair to $(N.x[:m], N.T[:d+1])$ and we generate a sequence of
other pairs $(x_i, T_i)$ from other sets. Since we do not know the minimal
Hamming distance from $X_{N.f}$ to other sets, we generate inputs $x_i$ by
gradually increasing $H(N.x[:m], x_i)$ as we generate more inputs. Namely, we
first generate all 1-bit mutations of $N.x[:m]$ (i.e., first $m \choose 1$
generated inputs), then all 2-bit mutations of $N.x[:m]$ (i.e., next $m \choose
2$ generated inputs), and so on.

Unfortunately, it turns out from our evaluation that performing more than 1-bit
mutations has negative impact on the overall performance of the \fizzer{}. In
fact, even 1-bit mutations already represent a considerable portions of all
inputs produced by the tool during the whole analysis. In order to deal with
the situation we implemented the following two approaches:
\begin{itemize}
    \item The evaluation also reviled that 1-bit mutations under-approximate the
        true set of sensitive bits a lot. Since we cannot generate higher bit
        mutations, we extended the detection of sensitive bits to byte
        boundaries, i.e., whenever a bit is detected as sensitive, then all bits
        in the same input byte are automatically marked as sensitive as well.
    \item Although the approach above increased the precision considerably, we
        also generate sequences of ``extreme'' bits - those with high Hamming
        distance from a randomly generated bits. For this we use the information
        about types in $N.t$:
        \begin{itemize}
            \item Bits corresponding to integer types we set to all zeros and
                also all to one.
            \item Bits corresponding to floating point types we set -1, 1, and
                to special values, like \code{INF}, \code{NAN}, \code{EPSILON}.
        \end{itemize}
        We also observed these ``extreme'' values provide a considerable chance
        to accidentally uncover ``special'' paths in the \target{}.
\end{itemize}

Since we detect sensitive bits w.r.t. the fixed pair $(N.x[:m], N.T[:d+1])$, we
can detect sensitive bits simultaneously for multiple nodes in $\pth{N}$.
Indeed, for each $0 \leq k \leq d$ we know $N.T[k].f$ and we also know the
number of bits we should consider, namely $8 \cdot N.T[k].nbytes$. Therefore,
for each generated input $x$, obtained from $X.x[:m]$ either by 1-bit mutation
or by the ``extreme'' values mutation, we obtain the corresponding trace $T$,
which we then map to nodes of the execution tree. Namely, if $K$ is the greatest
index such that for all $0 \leq k \leq K$ and $0 \leq l < K$ we have $T[k].id =
N.T[k].id$ and $T[k].direction = N.T[k].direction$, then we extend the mapping
of each $T[k]$ to $\pth{N}[k]$ by the sensitive bit(s) check:
\begin{itemize}
    \item 1-bit mutation: If $s$ is the index of the mutated bit, $s < 8 \cdot
        N.T[k].nbytes$, and $T[k].f \neq N.T[k].f$, then the bit at the
        index $s$ is sensitive in the node $\pth{N}[k]$ (and also all other bits
        in the same byte).
    \item ``extreme'' value mutation: The same procedure as above repeated for
        each bit index into the mutated value.
\end{itemize}

\paragraph{Notation} For each tree node we store the set $N.sbits$ of indices of
all sensitive bits detected by the sensitivity analysis.

\subsection{Bitshare analysis}\label{sec:bitshare}

Let us consider a node $N$ in the execution tree such that $N.sbits \neq
\emptyset$ and also a \code{Boolean} value $b$ such that $N.label[b] =
\code{NOT\_VISITED}$. The goal of the bitshare analysis is to find an input $x$
so that the obtained trace $T$ is mapped to $\pth{N}$, $|T| > dN + 1$ and
$T[dN].direction = b$.

The analysis looks for each node $M$ in the tree such that $M.id.uid =
N.id.uid$, $M.sbits \neq \emptyset$ and $M.label[b] \neq \code{NOT\_VISITED}$.
Observe that we intentionally ignore the calling context $M.id.ctx$. Although
the $\pth{N}$ and $\pth{M}$ represent different sequences of \code{Boolean}
instructions, they both pass through the instruction under question (possibly
even more than once). Since $M.x$ evaluated the instruction to $b$, then we
could try to somehow compose $N.x$ and $M.x$ so that the resulting input $x$
would produce a trace $T$ as described above.

The composition of $N.x$ and $M.x$ to $x$ is based on the sensitive bits
$N.sbits$ and $M.sbits$. First we initialize $x$ to be equal to $N.x$. Then we
build sorted
\footnote{Using the standard ``$<$'' order on the set of integers.}
sequences $I$ and $J$ of indices in $N.sbits$ and $M.sbits$, respectively. Now
for each $0 \leq i < min\{ |I|, |J| \}$ we set $x[I[i]] = M.x[J[i]]$.

Clearly, there are more ways how to use the sequences $I$ and $J$ for mapping
the sensitive bits of $M$ to to $x$. But we do not have information telling us
which is better. So, we use the most straightforward approach.

The described approach, of course, does not guarantee the obtained trace $T$ for
$x$ will be mapped to $\pth{N}$. But if it does, then there is reasonable chance
the instruction evaluates to $b$ (see the evaluation results).

\paragraph{NOTE:} Since the execution tree can be large, the analysis in fact
does not search the tree for all such nodes $M$. Instead, whenever any of the
two minimization analyses, started on some node $M$, is force terminated, i.e.,
the \code{Boolean} instruction was evaluated to the desired value $b$, then the
bitshare analysis is informed about that, meaning that it updates its map from
instruction unique IDs ($id.uid$ fields) and evaluation results ($direction$
fields) to values of sensitive bits of $M$. When the bitshare analysis is
started, then it uses input bits stored in its map. 

\subsection{Typed minimization analysis}\label{sec:typed_minimization}

The goal of the analysis is the same as of bitshare analysis (see the first
article in \rsec{sec:bitshare}). However, the analysis can be started for the
node $N$, only if each sensitive input bit $N.x[s]$, where $s \in N.sbits$,
belongs to a range of bits in $N.x$ associated with a type in $N.t$ such that
the type is none of \code{UNTYPED*} types (see \rsec{sec:fuzzing_loop}). The
reason for this requirement is that the analysis works on typed numerical
variables.

\begin{figure}
    \centering
    \includegraphics[width=10cm]{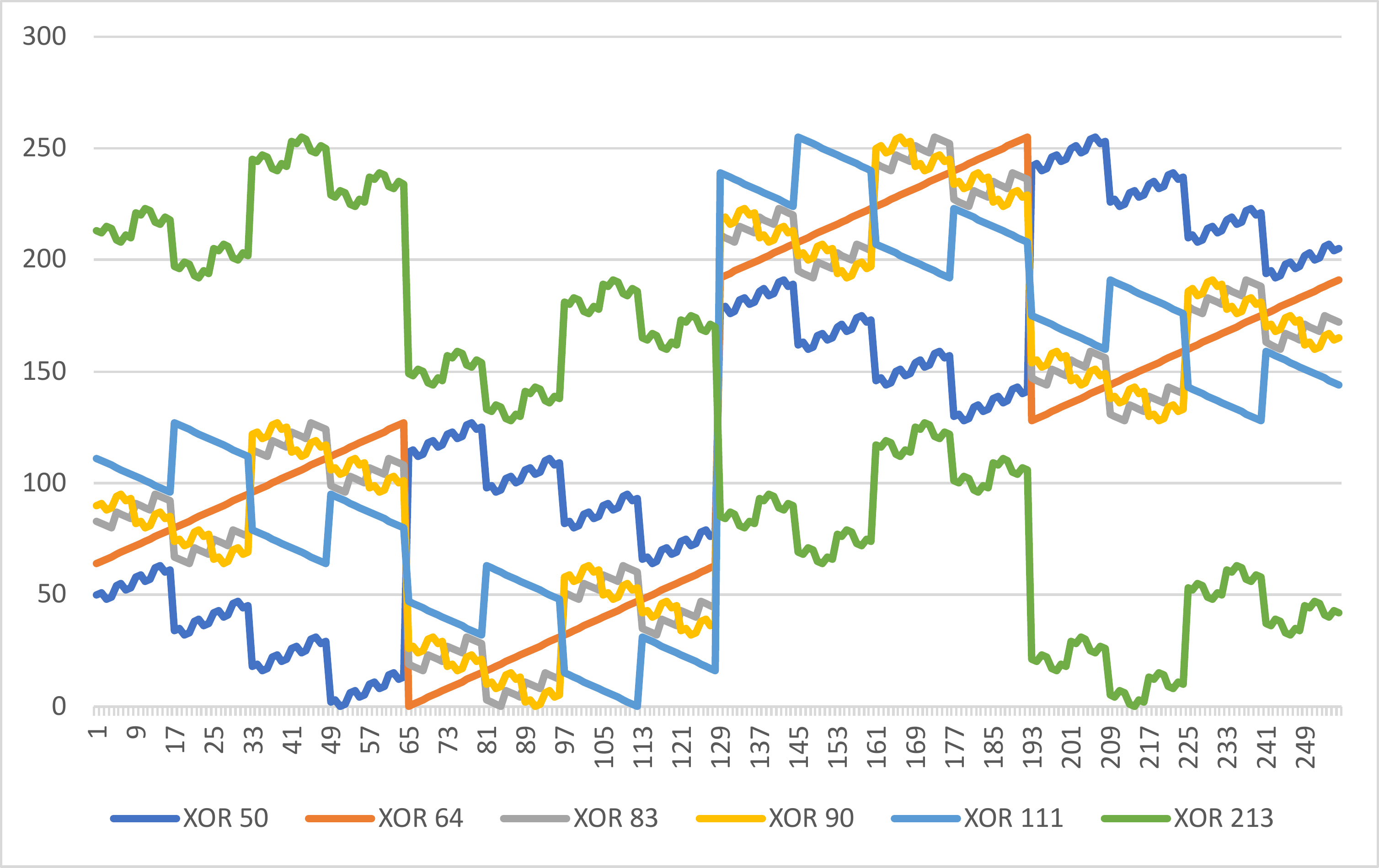}
    \caption{Plots of \code{x xor D} function for 8-bit variable
    \code{x} and few fixed constants \code{D}.}
    \label{pic:xor_hist}
\end{figure}

Another situation when this analysis is not used, if $N.xor$ is \code{true}.
When $xor$ instruction is used in a branching function, it then often has a lot
of local minima which are difficult to escape from (see \rfig{pic:xor_hist}).
Although the gradient descent is not effective for branching function with
\code{xor} in general, the version presented in \rsec{sec:minimization} performs
slightly better in more cases. Therefore, we leave the analysis of nodes with
$N.xor$ being \code{true} to the other algorithm.

\begin{figure}
    \centering
    \includegraphics[width=10cm]{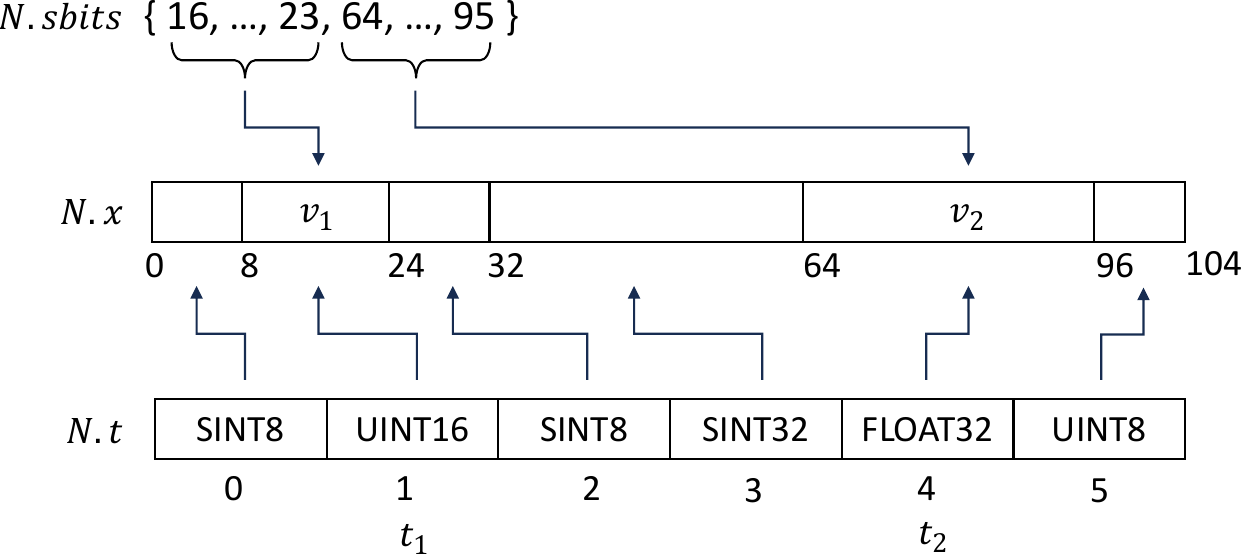}
    \caption{An example of typed numerical variables $v_1$ and $v_2$ with types
        $t_1$ and $t_2$.}
    \label{pic:gradient_descent_vars_example}
\end{figure}

The analysis thus starts by identifying typed numerical variables $\vec{v} =
(v_1, \ldots, v_m)$ in $N.x$ with types $\vec{t} = (t_1, \ldots, t_m)$ in $N.t$
using $N.sbits$. An example of this process is depicted in
\rfig{pic:gradient_descent_vars_example}. There we identify two variables, since
bit indices in $N.sbits$ points only to two regions associated with types in
$N.t$. Observe that not all bits of the variable $v_1$ are sensitive. That is
all right, because they are ignored in the construction of inputs. 

\begin{algorithm}
    \caption{Typed gradient descent}\label{alg:typed_gradient_descent}
    \begin{algorithmic}[1]
        \Loop\label{alg:typed_gradient_descent:outer_loop}
            \State $\vec{v} := $ generate next seed\label{alg:typed_gradient_descent:seed}
            \State $f(\vec{v}) := \code{ExecuteTarget}(\vec{v})$
            \If{$f(\vec{v})$ is not finite}{\textbf{~continue}}\EndIf
            \Loop\label{alg:typed_gradient_descent:inner_loop}
                \ForAll{$i = 1, \ldots, m$}\label{alg:typed_gradient_descent:partials}
                    \State Compute the smallest $\Delta v_i > 0$ s.t.
                            $v_i + \Delta v_i \neq v_i$.
                    \State $\nabla_i f(\vec{v}) := 
                            \frac{|\code{ExecuteTarget}(v_1, \ldots, v_i + \Delta v_i, \ldots, v_m)|
                                  - |f(\vec{v})|}{\Delta v_i}$
                    \If{$\nabla_i f(\vec{v})$ is finite}{
                        \code{lock}$[i] := $\code{false}
                    }\Else{
                        $\nabla_i f(\vec{v}) := 0$, \code{lock}$[i] := $\code{true}
                    }\EndIf
                \EndFor
                \State $success := $\code{false}
                \While{$||\nabla f(\vec{v})||^2$ is finite and for some $i$ we have $\code{lock}[i] = \code{false}$}
                    \label{alg:typed_gradient_descent:doing_a_step}
                    \State $\lambda := |f(\vec{v})| / ||\nabla f(\vec{v})||^2$
                        \label{alg:typed_gradient_descent:lambda}
                    \If{$\lambda$ is zero or not finite}{~\textbf{break}}\EndIf
                    \State $V' = \emptyset$
                    \ForAll{$e = 0, -1, 1, -2, 2, -3, 3$}
                        \label{alg:typed_gradient_descent:e_loop}
                        \State $\vec{v}' := \vec{v} - 10^e \lambda \nabla f(\vec{v})$
                            \label{alg:typed_gradient_descent:use_lambda}
                        \State $f(\vec{v}') := \code{ExecuteTarget}(\vec{v}')$
                        \State $V' := V' \cup \{ (\vec{v}', f(\vec{v}')) \}$
                    \EndFor
                    \State Let $(\vec{v}', f(\vec{v}')) \in V'$ be the pair with the smallest $|f(\vec{v}')|$
                    \If{$|f(\vec{v}')| < |f(\vec{v})|$}
                        \State $\vec{v} := \vec{v}'$, $f(\vec{v}) = f(\vec{v}')$, $success := \code{true}$
                        \State \textbf{break}
                    \Else\label{alg:typed_gradient_descent:lock_more_of_them}
                        \State $L := \{ 1/\nabla_i f(\vec{v})^2 ~|~ i = 1, \ldots, m$ and \code{lock}$[i] = \code{false}\}$
                        \State $l := min(L) + 0.6 * (max(L) - min(L))$
                        \ForAll{$i = 1, \ldots, m$ s.t. $\code{lock}[i] = false$}
                            \If{$\nabla_i f(\vec{v})^2 = 0$ or $1/ \nabla_i f(\vec{v})^2 < l$ or not finite}
                                \State $\nabla_i f(\vec{v}) := 0$, \code{lock}$[i] := $\code{true}
                            \EndIf
                        \EndFor
                        \If{no direction was locked in the loop above}{~\textbf{break}}\EndIf
                    \EndIf
                \EndWhile
                \If{$success = $\code{false}}{~\textbf{break}}\EndIf
            \EndLoop
        \EndLoop
    \end{algorithmic}
\end{algorithm}

Next follows the gradient descent of the unknown branching function $f(\vec{v})$
associated with the evaluation of the \code{Boolean} instruction, which
corresponds to the node $N$. The process is depicted in
\ralg{alg:typed_gradient_descent}.
We see that the computation happens in a seemingly infinite loop (see line
\ref{alg:typed_gradient_descent:outer_loop}). The algorithm terminates, when the
number of calls to \code{ExecuteTarget} exceed a certain limit
\footnote{The algorithm can also be force terminated any time from outside.}
, say $K$.
\footnote{In our implementation we use an empirically adjusted number $100\cdot|N.sbits|$.}
The check against the limit happens inside \code{ExecuteTarget}. If the limit is
exceeded the whole analysis is deactivated (meaning the search strategy is
finished). The function \code{ExecuteTarget} emulates the part of the fuzzing
loop, where the \target{} is executed for the passed input $\vec{v}$ and the
obtained trace $T$ is mapped to the node $N$. If the $T$ does actually not map
to $N$, then the function return $\infty$, representing a failure. Otherwise,
the function returns the value $T[dN].f$. The gradient descent algorithm cannot
work with infinite values. Therefore, if $T[dN].f$ is $\infty$, then it is also
considered as a failure.

In each iteration of the outer loop we first try to compute a seed input
$\vec{v}$ for which we want to get a valid (finite) $f(\vec{v})$. Once we
succeed we enter the inner loop at line
\ref{alg:typed_gradient_descent:inner_loop} where we perform the gradient
descent.

The process of seed generation depends on types in $\vec{t}$ and also on the
actual number of calls to \code{ExecuteTarget}. If the number of bits of a type
$t_i$ is smaller than 16, then we uniformly sample from the entire domain the
variable $v_i$ (i.e., from all possible values of the type $t_i$). For $t_i$
with the size 16 bits or more, we uniformly sample from a certain interval of
values of the domain. The bounds of the interval are functions of the number of
already performed calls, say $k$, to \code{ExecuteTarget}. Namely, for signed
integer type with $|t_i|$ bits the interval is $[-p, p]$, where $p = 2^{7 +
(|t_i|-9) k/K}$. For unsigned integer type the interval is $[0, p]$, where $p =
2^{7 + (|t_i|-8) k/K}$. And for floating type the interval is $[-p, p]$, where
$p = 2^{7 + (q-8) k/K}$ and  $q$ is $119$ for \code{float} and $115$ for
\code{double}. All numeric constants were adjusted empirically. The general idea
behind the process is to expand the sampled interval more and more as we
approach closer and closer to the limit $K$ of \target{} executions.

In the inner loop at line \ref{alg:typed_gradient_descent:inner_loop} we perform
the gradient descent from the seed input stored in $\vec{v}$. Each iteration of
the loop represents a single successful descent step, i.e., we have found a new
$\vec{v}$ such that $|f(\vec{v})|$ decreased.

In the loop at line \ref{alg:typed_gradient_descent:partials} we numerically
compute coordinates $\nabla_i f(\vec{v})$, one for each variable $v_i$, of the
gradient vector $\nabla f(\vec{v})$. Observe the coordinates are computed using
right differences, since $\Delta v_i > 0$. The computation of $\Delta v_i$ for
$t_i$ being an integer type is simple. We always choose $\Delta v_i = 1$. For
the floating point type $t_i$ we must take into account the value of $v_i$. For
example, if $v_i = 10^{20}$ and we choose $\Delta v_i = 1$, then we get $v_i +
\Delta v_i = v_i$, which is something we do not want.

For each coordinate $\nabla_i f(\vec{v})$ we also maintain \code{Boolean} flag
$\code{lock}[i]$ which can temporarily lock, i.e., disable, the coordinate from
the descent. We lock the coordinate if the value $\nabla_i f(\vec{v})$ is not
finite or if it considerably reduces the speed of the descent.

In the loop at line \ref{alg:typed_gradient_descent:doing_a_step} we use the
gradient vector $\nabla f(\vec{v})$ for finding a new input $\vec{v}'$ and the
corresponding value $f(\vec{v}')$ such that $|f(\vec{v}')| < |f(\vec{v})|$. We
perform the search till there is at least one gradient coordinate available for
use (i.e., not locked), and the magnitude of the gradient vector is finite.

The computation of the parameter $\lambda$ at line
\ref{alg:typed_gradient_descent:lambda} represents the core of the descent,
because we use it for computation of new input vectors $\vec{v}'$ (see line
\ref{alg:typed_gradient_descent:use_lambda}). We compute $\lambda$ under an
assumption that the branching function is linear around $\vec{v}$ so that we
can get to zero in single step. More precisely we want to compute the new input
$\vec{v}'$ as the intersection of the line
$$
    \begin{pmatrix} \vec{v} \\ 0 \end{pmatrix}
    -
    \lambda \begin{pmatrix} \nabla f(\vec{v}) \\ 0 \end{pmatrix}
$$
and a (hyper)plane
$$
    \begin{pmatrix} \vec{v} \\ |f(\vec{v})| \end{pmatrix}
    +
    t_1 \begin{pmatrix} \vec{e}_m^1  \\ \nabla_1 f(\vec{v}) \end{pmatrix}
    + \hdots +
    t_m \begin{pmatrix} \vec{e}_m^m \\ \nabla_m f(\vec{v}) \end{pmatrix}
$$
where $\vec{e}_m^i$ is the vector of the $i$-th coordinate axis in the
$m$-dimensional vector space. So, we solve for $\lambda$
$$
    \begin{pmatrix} \vec{v} \\ 0 \end{pmatrix}
    -
    \lambda \begin{pmatrix} \nabla f(\vec{v}) \\ 0 \end{pmatrix}
    =
    \begin{pmatrix} \vec{v} \\ |f(\vec{v})| \end{pmatrix}
    +
    t_1 \begin{pmatrix} \vec{e}_m^1  \\ \nabla_1 f(\vec{v}) \end{pmatrix}
    + \hdots +
    t_m \begin{pmatrix} \vec{e}_m^m \\ \nabla_m f(\vec{v}) \end{pmatrix}
$$

$$
    \begin{array}{rcl}
        -\lambda \nabla_1 f(\vec{v}) &= &t_1 \\
                                        &\vdots& \\
        -\lambda \nabla_m f(\vec{v}) &= &t_m \\
        0 &= &|f(\vec{v})| + t_1 \nabla_1 f(\vec{v}) + \cdots + t_m \nabla_m f(\vec{v})        
    \end{array}
$$
We can substitute variables $t_i$ to the last equation
$$
    \begin{array}{rl}
        0 =& |f(\vec{v})| + (-\lambda \nabla_1 f(\vec{v})) \nabla_1 f(\vec{v}) + \cdots + (-\lambda \nabla_m f(\vec{v})) \nabla_m f(\vec{v})\\
        0 =& |f(\vec{v})| - \lambda (\nabla_1 f(\vec{v})^2 + \cdots + \nabla_m f(\vec{v})^2)\\
        0 =& |f(\vec{v})| - \lambda ||\nabla f(\vec{v})||^2\\
        \lambda =& |f(\vec{v})| / ||\nabla f(\vec{v})||^2
    \end{array}
$$

In practice the branching function is not linear around $\vec{v}$, so we
generate several inputs from $\vec{v}$ in the opposite direction of $\nabla
f(\vec{v})$. That is done in the loop at line
\ref{alg:typed_gradient_descent:e_loop}. Observe that we generate inputs such
that the parameters $10^e$ range over seven orders of magnitude. So, we perform
smaller steps than $\lambda$ (up to 3 orders of magnitude) and larger steps than
$\lambda$ (also up to 3 orders of magnitude). This approach tackles two important
problems:
\begin{itemize}
    \item When the gradient descent converges the a neighborhood close to global
        minimum of the branching function the generated inputs sample that
        neighborhood.
    \item The gradient descent is more robust, meaning the generated input
        samples increase change of escaping from a local minima.
\end{itemize}

The code in the ``else'' branch (below the at line
\ref{alg:typed_gradient_descent:lock_more_of_them}) further improves and
robustness and also effectivity of the descent. If some $\nabla_i f(\vec{v})$ is
extremely large compare to other coordinates, then the vector $-\lambda \nabla
f(\vec{v})$ tends to change those other coordinates only negligibly. By locking
the coordinate with the extreme value we allow a descent in a new direction.

\subsection{Minimization analysis}\label{sec:minimization}

The goal of this minimization analysis is the same as of the typed minimization
(see \rsec{sec:typed_minimization}). Both analyses in fact apply the same kind
of algorithm -- the gradient descent. The key difference is that this analysis
does not use the information about types of bits in the input. So, the analysis
can be started for the node $N$, if some sensitive input bit $N.x[s]$, where $s
\in N.sbits$, belongs to a range of bits in $N.x$ associated with a type in
$N.t$ being some of \code{UNTYPED*} types (see \rsec{sec:fuzzing_loop}). This
analysis is also used, if $N.xor$ is \code{true}. That is for reasons we already
discussed in \rsec{sec:typed_minimization}. In summary, this analysis is applied
for nodes, for which the typed minimization either cannot work (missing
information about types), or when this analysis is expected to perform better
(\code{xor} instructions).

The minimization analysis applies the gradient descent algorithm. In contrast to
the gradient descent of the typed minimization analysis (see
\rsec{sec:typed_minimization}), here we consider each sensitive bit $N.x[s]$,
where $s \in N.sbits$, as an independent variable of the \code{Boolean} type.
So, we assume we have $m = |N.sbits|$ variables $\vec{v} = (v_1, \ldots, v_m)$,
where each $v_i$ can either be 0 or 1. This has, of course, an impact on the
structure and functionality of the algorithm. It is depicted at
\ralg{alg:binary_gradient_descent}.

\begin{algorithm}
    \caption{Binary gradient descent}\label{alg:binary_gradient_descent}
    \begin{algorithmic}[1]
        \State Generate a sequence $S$ of all seed inputs
        \ForAll{$\vec{v}_{seed}$ in $S$}
            \State $\vec{v} := \vec{v}_{seed}, f(\vec{v}) := |\code{ExecuteTarget}(\vec{v})|$
            \State $mag := [0, \ldots, 0]$ s.t. $|mag| = m$ ~~// where $m = |N.sbits|$
            \Loop\label{alg:binary_gradient_descent:loop}
                \ForAll{$i = 1, \ldots, m$}\label{alg:binary_gradient_descent:step_functions}
                    \State $f_i(\vec{v}) := |\code{ExecuteTarget}(v_1, \ldots, \neg v_i , \ldots, v_m)|$
                        \label{alg:binary_gradient_descent:f_i}
                    \State $mag[i-1] := \max\{ mag[i-1], |f_i(\vec{v}) - f(\vec{v})| \}$
                        \label{alg:binary_gradient_descent:mag}
                \EndFor
                \State $k = \arg \underset{i}{\min} \{ f_i(\vec{v}) ~|~ i=1,\ldots,m \}$
                    \label{alg:binary_gradient_descent:k}
                \If{$f_k(\vec{v}) < f(\vec{v})$}
                    \State $v_k := \neg v_k$, $f(\vec{v}) := f_k(\vec{v})$
                        \label{alg:binary_gradient_descent:move}
                    \State \textbf{continue}
                \EndIf
                \State Let $I$ be a permutation of $0,\ldots,m-1$ s.t. $\forall i,j.i \leq j \rightarrow mag[I[j]] \leq mag[I[i]]$\label{alg:binary_gradient_descent:permutation}
                \ForAll{$i = 1, \ldots, m$}\label{alg:binary_gradient_descent:step_functions2}
                    \State Let $\vec{v}_i$ be $\vec{v}$ with inverted bits at all indices $I[j]$ s.t. $j \geq i - 1$
                        \label{alg:binary_gradient_descent:v_i}
                    \State $f_i(\vec{v}_i) := |\code{ExecuteTarget}(\vec{v}_i)|$
                \EndFor
                \State $k = \arg \underset{i}{\min} \{ f_i(\vec{v}_i) ~|~ i=1,\ldots,m \}$
                \If{$f_k(\vec{v}_k) < f(\vec{v})$}
                    \State $\vec{v} := \vec{v}_k$, $f(\vec{v}) := f_k(\vec{v}')$
                        \label{alg:binary_gradient_descent:move2}
                    \State \textbf{continue}
                \EndIf
                \State \textbf{break}
            \EndLoop
        \EndFor
    \end{algorithmic}
\end{algorithm}

The algorithm starts by generating all seeds the algorithm may possibly use. We
define the count as a function of sensitive bits. Namely, we want to generate
about $m+1$ seeds. This count was established empirically. The goal is to sample
the set of all $2^m$ possible $m$-bit inputs uniformly. We can partition all
inputs into $m + 1$ classes $C_0, \ldots, C_m$ according to their Hamming
distance from the input $\vec{0} = [0, \ldots, 0]$. Observe that $|C_i| = {m
\choose i}$.
\footnote{Sizes of the classes thus form the row $m$ of the Pascal's
    triangle (rows being indexed from 0).}
For an uniform sampling we should take more samples from larger classes.
Fortunately, the number of samples we want to generate correlates with the
number of classes. So, we take one randomly chosen input from each class as a
seed (for $C_i$ we flip $i$ randomly chosen (yet different) bits in $\vec{0}$).

The rest of the algorithm operates with inputs of the size $m$ although the
actual size of the input is $|N.x| \geq m$. This is possible, because any
$m$-bit input $\vec{v}$ passed to \code{ExecuteTarget} is used together with
$N.x$ and $N.sbits$ to build the actual input $x$ for the \target{}. Namely, $x$
is first initialized to $N.x$ and then, if we assume the indices in $N.sbits$
are ordered by the standard ``$<$'', then for each $i = 1, \ldots, m$ we set
$x[N.sbits[i]]$ to $\vec{v}[i]$. Another important assumption about the
\code{ExecuteTarget} function always returns a finite floating point value. More
precisely, when the \server{} accepts a trace $T$ from the \target{} (executed
on the input $x$), then it returns $T[dN].f$ if the trace is mapped to $N$ and
$T[dN].f$ is finite. Otherwise, the maximal \code{double} value is returned.
Since the goal of the algorithm is to approach the global minimum of the
branching function, the maximal \code{double} value represents of the worst
possible outcome.

The minimization algorithm takes generated seeds sequentially one by one and for
each it tries to apply the binary gradient descent to approach the global
minimum in a hope of inverting the evaluation result of the \code{Boolean}
instruction associated with the node $N$ along the way. The descent is inside
the loop at line \ref{alg:binary_gradient_descent:loop}. There we first sample
the branching function around $\vec{v}$ for 1-bit mutations, for each bit index
$i$ one mutation, in order to obtain the absolute values of the corresponding
branching function values $|f_i(\vec{v})|$. That is done in the loop at line
\ref{alg:binary_gradient_descent:step_functions}. Observe that we actually do
not compute partial derivatives of the branching function. That is because we
perform the step only in one of $m$ coordinates of the gradient, i.e., in the
coordinate $k$ such that $f_k(\vec{v})$ is the smallest.
\footnote{The computation of $\nabla_k f(\vec{v}) = (f_k(\vec{v}) -
    f(\vec{v})) / 1$ would be useless, because the $f(\vec{v}) := f(\vec{v}) +
    1 \cdot \nabla_k f(\vec{v}) = f_k(\vec{v})$.}
This index $k$ is computed at line \ref{alg:binary_gradient_descent:k}. If we
further have $f_k(\vec{v}) < f(\vec{v})$, then we move in the direction of the
coordinate $k$ (see line \ref{alg:binary_gradient_descent:move}) and we continue
to the next gradient step. The decision for modifying the standard version of
the gradient descent so that we step only in direction (coordinate) is based on
our practical experience with the algorithm -- the single coordinate version is
more robust, i.e., it has a higher success rate of escaping from a local
minimum, in a price of decreased effectivity. Since we use this algorithm mostly
for branching functions with lots of local minima (like \code{xor} function, see
\rfig{pic:xor_hist}), the robustness is more valuable than convergence speed.

Observe that the code at lines
\ref{alg:binary_gradient_descent:step_functions2}--\ref{alg:binary_gradient_descent:move2}
look similar to the binary gradient step described above (lines
\ref{alg:binary_gradient_descent:step_functions}--\ref{alg:binary_gradient_descent:move}).
There is one key difference though. In the construction of the mutated inputs
$\vec{v}_i$ at line \ref{alg:binary_gradient_descent:v_i}, in contrast to line
\ref{alg:binary_gradient_descent:f_i}, more than one bit can be mutated.
These multi-bit mutations are targeted to situations when some sensitive
bits $N.sbits$ collectively behave as an integer. It is easy to show that
convergence from one integer to another using only single bit mutations can
get stuck in a local minimum. Let us consider this program
\begin{verbatim}
    char x = __VERIFIER_nondet_char() & 15;
    bool bi = x == 4;
\end{verbatim}
We clearly have four sensitive bits $N.sbits = \{4,5,6,7\}$ and our branching
function is $f(x) = x - 4$. Observe that for $\vec{v} = (0,0,1,1)$ we have
$|f(\vec{v})| = 1$ and there is no single-bit mutation $\vec{v}'$ of $\vec{v}$
for which $|f(\vec{v}') < |f(\vec{v})|$. So, $\vec{v}$ is a local minimum.
However, we can escape from it by mutating the last 3 bits simultaneously, i.e.,
we get $\vec{v}' = (0,1,0,0)$ and $f(\vec{v}') = 0$. Observe also, that we can
obtain $(0,1,0,0)$, if we trait $\vec{v}$ as an integer and we added 1 to it.
So, the idea behind our multi-bit mutations is to increment $\vec{v}$ by 1 in a
hope to escape the described local minimum, if we happened to get stuck there.
However, we need to know the importance of the sensitive bits in the integer
value for the implementation of the incrementation. Let us insert the following
line in between the two lines of code above
\begin{verbatim}
    x = ((x & 1) << 3) | (x & 6) | (x & 8) >> 3;
\end{verbatim}
This line swaps the bits at indices 4 and 7. Clearly, the desired input we seek
is now $\vec{v}' = (0,1,0,0)$, which means that we need to mutate a different 3
bits then previously.

We thus always need to detect the importance of bits. We do so in the binary
gradient descent, at line \ref{alg:binary_gradient_descent:mag}, by computing
elements of the sequence $mag$. An element $mag[i]$ stores the maximum
difference between values $f_i(\vec{v})$ and $f(\vec{v})$ computed during all
gradient steps from a given seed. The higher the value $mag[i]$ the higher
importance of the the sensitive bit $i$. The permutation $I$ then represents the
order of sensitive bits in the decreasing importance.

In the example above we guessed which 3 bits should be mutated to escape from
the local minimum. Unfortunately, in general we do not know what sub-sequence of
$m$ sensitive bit should be mutated. Therefore we try all $m$ of them (see the
loop at line \ref{alg:binary_gradient_descent:step_functions2}).

\section{Selection of input generator}\label{sec:generator_selection}

Whenever none of the four input generation analysis is active there must be some
of them selected and activated. Each of these analyses operates with some node
$N$ of the execution. The node is passed to the analysis as an argument of the
activation. Therefore, the selection process of an analysis to activate starts
with a search for a node $N$ in the tree. The analysis is then selected based on
node's properties.

Only nodes corresponding to \code{Boolean} instructions which has not been
covered yet are considered in the search. We update the information about
coverage of \code{Boolean} instructions (and corresponding nodes) in each
iteration of the fuzzing loop; that happens during the process of mapping
accepted execution traces to the execution tree.

The node selection process works with the following properties of nodes:
\begin{itemize}
    \item A node $N$ is directly input dependent (DID), iff $N.sa \wedge N.sbits
        \neq \emptyset$.
    \item A node $N$ is indirectly input dependent (IID), iff $N.sa \wedge
        N.sbits = \emptyset$.
    \item A node $N$ is open iff there is a \code{Boolean} value $b$ s.t.
        $N.labels[b] = \code{NOT\_VISITED}$ and $\neg N.sa \vee (N.sbits \neq
        \emptyset \wedge (\neg N.ba \vee \neg N.ma))$
    \item A node $N$ is closed, iff it is not open and for both
        \code{Boolean} values $b$ either $N.labels[b] \neq \code{VISITED}$ or
        $N.succ[b]$ is closed.
\end{itemize}

Before we continue further let us look at the following observations:
\begin{itemize}
    \item The check for a node being closed depends on successor node(s) being
        closed (if there is some). It means that first closed nodes are leaves
        of the tree, then their parents, and so on up to the root node.

        Note: We update closed state of nodes whenever an analysis is
        deactivated. The update starts from the node the analysis was started
        with and continues towards the root of the execution tree.
    \item A node cannot be DID and IID in the same time, but it can be neither
        DID nor IID. The same we can say also for open and closed properties.
\end{itemize}

We first search for the node amongst primary coverage targets (discussed later
in \rsec{sec:primary_targets}). If the search fails, then we continue by a Monte
Carlo search from some IID pivot (also discussed later in
\rsec{sec:monte_carlo}). If this search fails as well, then the analysis cannot
make any further progress and the fuzzing loop terminates.
\footnote{There are circumstances under which the termination of the fuzzing
loop can be resumed, meaning that we are able to make some nodes in the tree to
be primary targets. We discuss details of this later in
\rsec{sec:recovery_from_termination}.}
However, if some of the two searches succeeds, then we obtain the winning node
$N$ and we proceed with it to the selection of the analysis to be activated.
This process is depicted in \ralg{alg:select_analysis}. But before we look at it
we introduce two notations.

\paragraph{Notation} We allow to define a sequence in a ``set'' style, i.e., $[
f(v) ~|~ v = v_1, \ldots, v_n ]$ is the sequence $[ f(v_1), \ldots, f(v_n)]$.

\paragraph{Notation} For each node $N$ in the execution tree we introduce an
integer field $N.height$ which is updated for every execution trace $T$ mapped
to $N$ to a value $\max\{ N.height, |T| \}$. When the node $N$ is created and
inserted to the tree, then the field is initialized to $|T|$ (recall that a node
can be inserted to the tree only when some trace is mapped to the tree). Observe
the field actually stores the maximal from depths of all nodes in its
sub-tree(s), i.e., the value $\max\{ dM ~|~ N \in \pth{M} \}$.\\

\begin{algorithm}
    \caption{Select analysis for activation}\label{alg:select_analysis}
    \begin{algorithmic}[1]
        \State Selected a primary coverage target $N$ (see \rsec{sec:primary_targets})
        \If{$N = \code{null}$}{~Selected $N$ by the Monte Carlo method (see \rsec{sec:monte_carlo})} \EndIf
        \If{$N = \code{null}$}{~Terminate the fuzzing loop.} \EndIf
        \If{$\neg N.sa$}
            \Loop\label{alg:select_analysis:loop}
                \State $succ := [ N.succ[b] ~|~ b = \code{false}, \code{true} ]$
                \State $dir := [ succ[i] \neq \code{null} \wedge succ[i].nbytes = N.nbytes ~|~ i = 0,1 ]$
                \If{$dir[0] \wedge dir[1]$}
                    \State $N := succ[succ[0].height \geq succ[1].height ~\code{?}~ 0 ~\code{:}~ 1]$
                \ElsIf{~$dir[0]$}
                    \State $N := succ[0]$
                \ElsIf{~$dir[1]$}
                    \State $N := succ[1]$
                \Else
                    \State \code{break}
                \EndIf
            \EndLoop
            \State Select the sensitivity analysis with the node $N$.
        \ElsIf{$\neg N.ba$}
            \State Select the bitshare analysis with the node $N$.
        \ElsIf{$\neg N.xor$ and $N.sbits$ correspond to numerical variables of known types}
            \State Select the typed minimization analysis with the node $N$.
        \Else
            \State Select the minimization analysis with the node $N$.
        \EndIf
    \end{algorithmic}
\end{algorithm}

We are ready to discuss \ralg{alg:select_analysis}. In first two lines we try to
find a node $N$ in the tree to be used for the selection of the analysis (to be
then activated with the node). Details are discussed in
\rsec{sec:primary_targets} and \rsec{sec:monte_carlo}. If the selection of the
node $N$ fails, then we terminate the fuzzing loop, i.e., we terminate the
entire fuzzing process. Otherwise, we may proceed to the analysis selection.

If the sensitivity analysis has not been applied to $N$ yet, it is selected
(with possibly a node in $N$'s subtree; we discuss details below), because it
will compute the sensitive bits necessary for other analyses. Otherwise, we
attempt to select the bitshare analysis, because it is fast (it basically
retrieves inputs from the cache) and also effective. If the previous two
analyses are not available, then we select one of the two minimization analyses.
If the conditions for activation of the typed minimization are satisfied, then
we select it, because it performs better under these conditions. Otherwise, the
conditions are such that the minimization analysis is expected to perform
better, and so it is selected. Observe that although we do not check for $N.ma$
when choosing between the two minimization analyses, we are sure that $N.ma$ is
\code{false}, because the selected node $N$ is open. 

It remains to explain the purpose of the loop in the algorithm. Recall that
given a node $N$, the sensitivity analysis may compute or update the sensitive
bits of any node in $\pth{N}$. If we activate the analysis with some node in
$N's$ subtree (if there is any), then $N.sbits$ will still be computed and the
analysis will in addition compute $N.sbits$ of more nodes. So, it looks like we
achieve the best effectivity, if we start the analysis in a leaf node in $N's$
subtree at the highest depth. There is a catch however. Nodes below $N$ may
correspond to more input bytes, i.e., their field $nbytes$ can be greater than
$N.nbytes$. If such node is chosen, then the analysis will check for sensitivity
of bits at indices $\geq 8 \cdot N.nbytes$, which definitely cannot be sensitive
bits of $N$. So, we thus actually could decrease effectivity, especially if
$nbytes$ of the leaf node is much larger than $N.nbytes$. Also, if the node
selection algorithm wanted to apply the sensitivity for a longer input, then it
would selected some node below $N$ in the first place. Therefore, in order to be
sure we maximize the effectivity, we must consider only those nodes below $N$
with the same value in the field $nbytes$. But from all these nodes we may still
choose any with the maximal depth in the tree. The search for such node is
implemented in the loop in \ralg{alg:select_analysis}. Observe that we use the
field $height$ to navigate towards a leaf at the highest depth.

\subsection{Searching in primary coverage targets}\label{sec:primary_targets}

A primary target is a node appearing in any of the following
\begin{itemize}
    \item Loop heads: A set $\mathcal{H}$ of nodes. Each its node $N$
        correspond to an execution of a \code{Boolean} instruction representing
        the head of some loop along an execution trace mapped either to the node
        or some node in its subtree. Since these nodes represent borders between
        iterations of loops, execution traces mapped the yet not visited
        successor of $N$ may improve coverage of many nodes. Of course, a loop
        can be iterated many time, so we must computed which of all iterations
        are actually important for the overall effectivity of the \fizzer. We
        discuss details below.
    \item Sensitive: An ordered set $\mathcal{S}$ of open nodes $N$ such that
        $N$ was only processed by the sensitivity analysis, i.e., it was
        prepared for other input generation analyses (see
        \rsec{sec:input_generators}), but none of those has been activated with
        $N$ yet. The order of nodes has an impact on effectivity of the overall
        performance. We discuss it later.
    \item Untouched: An ordered set $\mathcal{U}$ of open nodes $N$ such that
        $N.sa$ is \code{false} and further $N.id$ is not the location of any IID
        pivot (see \rsec{sec:monte_carlo}). It means that $N$ can be processed
        by the four input generation analyses (see \rsec{sec:input_generators}),
        but it has not been ``touched'' by any of them yet. The order of nodes
        is given by the same relation as the one on the set of sensitive
        targets.
    \item IID twins: A sequence $\mathcal{T}$ of open nodes $N$ such that $N.sa$ is
        \code{false} and there is an IID pivot $M$ (see \rsec{sec:monte_carlo})
        such that $N.id = M.id \wedge |N.f(x)| < |M.f(x)|$. It means that $N$
        can be but has not been processed by any of the four input generation
        analyses (see \rsec{sec:input_generators}) yet. It also represents the
        same uncovered \code{Boolean} instruction as the node $M$. So, $N$ is a
        ``twin'' of $M$. Moreover, from the comparison of values of the
        branching function $f(x)$ the node $N$ is closer to the global minimum
        and so it has higher potential for covering the instruction than $M$.
        Therefore, even if an activation of the sensitivity analysis with $N$
        would result in $N.sbits = \emptyset$ (which is quite likely to happen),
        the node could still be valuable for the search discussed in
        \rsec{sec:monte_carlo}.
\end{itemize}

We insert nodes into $\mathcal{S}, \mathcal{U}, \mathcal{T}$ during the process
of mapping of each accepted execution trace $T$ to the execution tree. Namely,
for each node $N$ inserted to the tree we consider the insertion of $N$ to each
of them. The insertion of nodes into $\mathcal{H}$ is more complicated. It
happens during the actual node selection process. We discuss it later. We
further prune contents of all $\mathcal{H}, \mathcal{S}, \mathcal{U},
\mathcal{T}$ every time the active input generation analysis becomes
deactivated. We erase all those nodes which do not satisfy the criteria we
defined above.

\begin{algorithm}
    \caption{Select a primary target node}\label{alg:select_primary_node}
    \begin{algorithmic}[1]
        \If{$\mathcal{H} \neq \emptyset$}\label{alg:select_primary_node:loop_heads}
            \State Extract any node from $\mathcal{H}$ and return it.
        \ElsIf{$\mathcal{S} \neq \emptyset$}
            \State Let $N$ be the smallest node of the ordered set $\mathcal{S}$.
            \If{Loop heads has not been detected along $\pth{N}$}
                \State Detect loop heads along $\pth{N}$
                \State \textbf{goto} line \ref{alg:select_primary_node:loop_heads}.
            \Else
                \State Extract $N$ from $\mathcal{S}$ and return it.
            \EndIf
        \ElsIf{$\mathcal{U} \neq \emptyset$}
            \State Let $N$ be the smallest node of the ordered set $\mathcal{U}$.
            \If{Loop heads has not been detected along $\pth{N}$}
                \State Detect loop heads along $\pth{N}$
                \State \textbf{goto} line \ref{alg:select_primary_node:loop_heads}.
            \Else
                \State Extract $N$ from $\mathcal{U}$ and return it.
            \EndIf
        \ElsIf{$|\mathcal{T}| > 0$}
            \State $N := \mathcal{T}[0]$
            \If{Loop heads has not been detected along $\pth{N}$}
                \State Detect loop heads along $\pth{N}$
                \State \textbf{goto} line \ref{alg:select_primary_node:loop_heads}.
            \Else
                \State $\mathcal{T} := \mathcal{T}[1:]$, \textbf{return} $N$
            \EndIf
        \Else ~\textbf{return} \code{null}
        \EndIf
    \end{algorithmic}
\end{algorithm}

The process of selection a primary target node is depicted in
\ralg{alg:select_primary_node}. The procedure is straightforward. We take
$\mathcal{H}, \mathcal{S}, \mathcal{U}, \mathcal{T}$ in this exact order and we
look for the first one not being empty. If all are empty, no primary target can
be selected and we return \code{null}. Otherwise, we extract some node from the
non-empty set and we return it. From $\mathcal{H}, \mathcal{T}$ we choose the
node randomly.
\footnote{Although we in fact always take the first element of $\mathcal{T}$,
the order in which the nodes arrive, and which we always push to the end of
$\mathcal{T}$, is random (not important for us).}
From the ordered sets $\mathcal{S}, \mathcal{U}$ we use the order to take
the smallest node.

Observe that the selection of node from $\mathcal{S}, \mathcal{U}, \mathcal{T}$
can be interrupted, if the chosen node $N$ has not been considered in the loop
detection yet. In which case we perform the detection (which can make
$\mathcal{H}$ non-empty) and then basically restart the selection process (we
return to the first line). There is the following reason for this
implementation. The count of paths $\pth{N}$ in the tree, for any node $N$, can
be large. We thus cannot perform the computation for all nodes, since that would
have a serious negative impact on the overall performance of the \server{}. But
we know that we surely want to process the node selected from $\mathcal{S},
\mathcal{U}, \mathcal{T}$, i.e., we surely want to consider loop heads on the
path to that node. 

\subsubsection{Detection of loop heads}

The \server{} has no information about control flow structures, like branchings
or loops, in the \target{}. Therefore, when we speak about loops and loop heads,
we actually consider only repetitions of \code{Boolean} instructions along paths
in the execution tree.
\footnote{Nevertheless, a repetition of a \code{Boolean} instruction often means
that it in fact appears inside an actual loop inside the \target{}. }
We compute loop heads, to be inserted to $\mathcal{H}$, in three steps.
First we must select a node $N$ in the tree. We already know that it is a node
selected from any of $\mathcal{S}, \mathcal{U}, \mathcal{T}$.
In the second step we detect all loop heads along the path $\pth{N}$ using
\ralg{alg:detect_loops}. The algorithm is in fact more general, because it is
also used in \rsec{sec:monte_carlo}. So, we will describe it completely.

\begin{algorithm}
    \caption{Detect loops}\label{alg:detect_loops}
    \begin{algorithmic}[1]
        \State $loops := []$, $heads2bodies := \emptyset$
        \State $stack := []$, $lookup := \emptyset$
        \ForAll{$i = |\pth{N}| - 1, \ldots, 0$}
            \State $j := \min\{ i + 1, |\pth{N}| - 1 \}$
            \If{$\pth{N}[i].id \not\in Dom(lookup)$}\label{alg:detect_loops:new_id}
                \State $lookup[\pth{N}[i].id] := |stack|$
                \State $stack := stack + [(\pth{N}[i], \pth{N}[j], 0)]$
            \Else
                \State $k := lookup[\pth{N}[i].id]$
                \If{$stack[k].index = 0$}
                    \State $stack[k].index := |loops|$
                    \State $loops := loops + [(\pth{N}[i], stack[k].X, stack[k].S)]$
                \Else
                    \State $loops[stack[k].index].E := \pth{N}[i]$
                \EndIf
                \While{$|stack| > k + 1$}\label{alg:detect_loops:cleanup}
                    \State Insert $stack[-1].X.id$ to $heads2bodies[stack[k].X.id]$
                    \State Erase $stack[-1].X.id$ from $lookup$
                    \State Erase the last element from $stack$
                \EndWhile
            \EndIf
        \EndFor
        \ForAll{triples $L$ in $loops$}\label{alg:detect_loops:postprocess}
            \While{$L.E.parent \neq \code{null} \wedge (L.E.parent.id = L[1].id \vee L.E.parent.id \in heads2bodies[L.X.id])$}
                \State $L.E := L.E.parent$
            \EndWhile
        \EndFor
        \State \textbf{return} $loops, heads2bodies$
    \end{algorithmic}
\end{algorithm}

\paragraph{Loop detection}
The algorithm computes a sequence $loops$ and a map $heads2bodies$. An elements
of $loops$ is a triple $(E, X, S)$, called a loop boundary, where $E$ is the
tree node from which we enter to the loop, $X$ is the node from which we exit
from the loop, and $S$ is the successor of $X$ in $\pth{N}$. The map
$heads2bodies$ maps IDs of \code{Boolean} instruction detected as loop heads to
a set of IDs of all \code{Boolean} instructions representing the body of the
loop.

We compute both results by processing the path $\pth{N}$ backwards.
\footnote{We can exit from a loop only from the loop-head \code{Boolean}
instructions, but the loop does not have to start with it. Backward traversal
thus allows for easier detection of the loop heads.}
During this traversal we build a stack $stack$, where we stack the first
occurrences of \code{Boolean} instructions (their $id$s). An element of the
$stack$ is a triple $(X, S, index)$, where $X$ is the node from which we exit
from the loop, $S$ is the successor of $X$ in $\pth{N}$, and $index$ is the
index of the corresponding element in the sequence $loops$. We use a map
$lookup$ for mapping the first occurrences of \code{Boolean} instructions (their
$id$s) to indices the of the corresponding records in $stack$. Observe that we
check for the first occurrences at line \ref{alg:detect_loops:new_id}. In the
case of the first occurrence we extend both $stack$ and $lookup$ map. Otherwise,
we query the $lookup$ map to get the index $k$ of the triple in $stack$
representing the first occurrence of $\pth{N}[i].id$. The case when
$stack[k].index = 0$ identifies the first repetition of the \code{Boolean}
instruction with ID $\pth{N}[i].id$ along $\pth{N}$ (by going backwards).
Therefore, this is the first evidence that we are in a loop, and so we record
the loop in the sequence $loops$. Otherwise, this is some other iteration of the
loop. So we only move the entry to the loop to the current node $\pth{N}[i]$.
The loop at line \ref{alg:detect_loops:cleanup} erases everything from $stack$
and $lookup$ what was recorded since the first occurrence of the instruction,
which is the record at index $k$ in $stack$. Note that erased records represent
\code{Boolean} instructions forming the body of the loop. Therefore, we insert
all IDs of \code{Boolean} instructions corresponding to all erased records to
map $heads2bodies$.

The loop at line \ref{alg:detect_loops:postprocess} performs a postprocessing of
loop entries of all recorded loops. We basically do not want the entry and exit
nodes (instructions) be the same and we also do not want the entry to be in the
loop body. We resolve such situations by moving the entry towards the root node
of the execution tree.

\begin{algorithm}
    \caption{Detect loop heads}\label{alg:detect_loop_heads}
    \begin{algorithmic}[1]
        \State $W := \{ (2^i ,\emptyset) ~|~ i = 0, \ldots, 10 \}$
        \ForAll{$i = 0, \ldots, |\pth{N}| - 1$ s.t. $\pth{N}[i]$ is open and $\pth{N}[i].id \in Dom(heads2bodies)$}
            \State Let $w \in Dom(W)$ be s.t. $\forall w' \in Dom(W)~.~|\pth{N}[i].nbytes - w| \leq |\pth{N}[i].nbytes - w'|$
            \State Insert $\pth{N}[i]$ to the ordered set $W[w]$.
        \EndFor
        \ForAll{$H \in Rng(W)$}
            \State Insert the smallest element (node) in $H$ into $\mathcal{H}$.
        \EndFor
    \end{algorithmic}
\end{algorithm}

Now we are back at the detection of loop head for the set $\mathcal{H}$. We only
need the domain of the map $heads2bodies$ obtained from \ralg{alg:detect_loops}.
The computation of the loop head using $Dom(heads2bodies)$ is depicted in
\ralg{alg:detect_loop_heads}. In the first loop we collect all open loop head
nodes along $\pth{N}$. However, instead of inserting them all directly to
$\mathcal{H}$, we actually group them, according to the number of input bytes
read along the path up to them.

The reason for that comes from evaluations, where we observed that overall
performance of the tool is highly sensitive to the selection of loop heads. Any
of the following two serious performance issues may occur, if we do not filter
the loop heads (e.g., as we do in \ralg{alg:detect_loop_heads}):
\begin{itemize}
    \item Each loop head corresponds to a certain iteration of some loop. The
        number of loop iterations can be large. So, our analysis can easily get
        ineffective because of processing of just a lots of loop heads.
    \item More input bytes may be read or processed with the increasing count of
        loop iterations. Effectivity of all four input generation analyses
        depend on input size. So, a lot of effort can be spent just on the
        detection of sensitive bits by the sensitivity analysis, leading to
        a serious performance decrease.
\end{itemize}
Our approach to the issues is to keep both the count of loop heads and also the
number of processed input bytes in reasonable bounds. Therefore, we group all
loop heads into just 11 classes based on the number of input bytes. We use the
map $W$ for the grouping. $Dom(W)$ define classes of input size and $Rng(W)$ are
ordered sets of open loop head nodes $N[i]$. The exponential function $2^i$ allows
for more refine grouping for small input size and coarse grouping for large input size.
For example, it allows distinguishing between input sizes 4 and 8, while ignoring
the difference between the sizes 1000 and 1004.

Once the map $W$ is filled in, then we insert only one representative node from
each group into $\mathcal{H}$, namely the smallest representative. Given $H \in
Rng(W)$, then nodes in $H$ is ordered using the following strict order: Let $P,Q
\in H$. Then, $P < Q$, iff 

$$P.nbytes < Q.nbytes \vee (P.nbytes = Q.nbytes \wedge dP < dQ)$$.

\subsubsection{Order on the sets of sensitive and untouched targets}

The overall effectivity of the analysis not only depends on what nodes are
selected, but also when. Based on result of our evaluations, we established the
following strict weak order the sets $\mathcal{S}, \mathcal{U}$: Let $P,Q$ be
nodes in either $\mathcal{S}$ or $\mathcal{U}$ and $max\_bytes$ be the greatest
value of the field $nbytes$ of all nodes in the execution tree. Then, $P < Q$,
iff \ralg{alg:order_of_S_U} returns \code{true}.

\begin{algorithm}
    \caption{Strict weak ordering of $\mathcal{S}, \mathcal{U}$}\label{alg:order_of_S_U}
    \begin{algorithmic}[1]
        \If{$P.sa \wedge \neg Q.sa$}{~\textbf{return} \code{true}} \EndIf
        \If{$\neg P.sa \wedge Q.sa$}{~\textbf{return} \code{false}} \EndIf
        \If{$|P.sbits| < |Q.sbits|$}{~\textbf{return} \code{true}} \EndIf
        \If{$|P.sbits| > |Q.sbits|$}{~\textbf{return} \code{false}} \EndIf
        \State $W := [ 2^0, \ldots , 2^{10} ]$
        \State $p := \arg \underset{i}{\min}\{ | P.nbytes - W[i] | ~|~ i = 0, \ldots, |W|-1 \}$
        \State $q := \arg \underset{i}{\min}\{ | Q.nbytes - W[i] | ~|~ i = 0, \ldots, |W|-1 \}$
        \State $m := \arg \underset{i}{\min}\{ | max\_bytes/2 - W[i] | ~|~ i = 0, \ldots, |W|-1 \}$
        \If{$|W[m] - W[p]| < |W[m] - W[q]|$}{~\textbf{return} \code{true}} \EndIf \label{alg:order_of_S_U:go_to_center1}
        \If{$|W[m] - W[p]| > |W[m] - W[q]|$}{~\textbf{return} \code{false}} \EndIf \label{alg:order_of_S_U:go_to_center2}
        \If{$P.nbytes < Q.nbytes$}{~\textbf{return} \code{true}} \EndIf
        \If{$P.nbytes > Q.nbytes$}{~\textbf{return} \code{false}} \EndIf
        \If{$dP < dQ$}{~\textbf{return} \code{true}} \EndIf
        \If{$dP > dQ$}{~\textbf{return} \code{false}} \EndIf
        \State \textbf{return} $P.height > Q.height$
    \end{algorithmic}
\end{algorithm}

The purpose of the lines \ref{alg:order_of_S_U:go_to_center1} and
\ref{alg:order_of_S_U:go_to_center2} is to prefer nodes whose field $nbytes$ is
close to the half of the maximal number input bytes read so far.

\subsection{Monte Carlo search from IID pivot}\label{sec:monte_carlo}

The ultimate goal here is to cover those \code{Boolean} instructions whose
corresponding nodes in the execution tree all have empty set of sensitive bits.
We denote these nodes as IID nodes. Given an IID node $N$, we cannot activate
any of the four input generation analyses (see \rsec{sec:input_generators}). The
actual goal here is to search for such node $M$ in the tree, with which some of
the input generation analyses can be activated and the success of the analysis
(i.e., we get the opposite result from the evaluation of the \code{Boolean}
instruction) would get us closer to the coverage of $N$ (i.e., of the
\code{Boolean} instructions corresponding to $N$). By ``get us closer'' we
actually mean that once the analysis is done (deactivated), there may appear a
node $N'$ in a subtree of $M$ such that $N'.id = N.id \wedge |N'.f(x)| <
|N.f(x)|$. 

Before we explain, how we search for a node $M$, we first need to know what IID
nodes should actually be covered. We call them IID pivots and we detect them in
the execution tree whenever the sensitivity analysis is (forcefully)
deactivated. Each node changed by the analysis which is also uncovered IID node
(see \rsec{sec:primary_targets}) is a new IID pivot. Recall the sensitivity
analysis of a node $N$ may actually change sensitive bits of any node in
$\pth{N}$. A node stays as an IID pivot until it is covered.

We start our search by selecting an IID pivot which we would like to cover. This
is done in two steps. First we partition the set of all pivots by the field
$id$. That makes sense, because there can be several IID pivots in the tree
corresponding to the same \code{Boolean} instruction. In this step we just want
to decide, which of the \code{Boolean} instruction we focus on. Since we are not
aware of a meaningful information for ordering the instructions, we choose a
partition class $C$ of IID pivots randomly, using the uniform distribution. In
the second step we select a representative pivot from $C$. In contrast to the
previous step we have an information (inferred from our evaluations) to build a
strict weak order on $C$. It is depicted in \ralg{alg:order_of_IID_partition},
where $P,Q \in C$. Surprisingly, our evaluation shows that instead of always
choosing the smallest pivot in $C$, it is often more effective to actually
choose the representative pivot randomly, using a distribution biased towards
smaller elements in $C$. Namely, if we consider $C$ as an ordered sequence of
pivots, then the probability of choosing $C[i]$, where $0 \leq i < |C|$, is $p_i
= \frac{3}{4}(1 - p_{i-1}), p_0 = \frac{3}{4}$. The values $p_i$ decay
exponentially with increasing $i$.

\begin{algorithm}
    \caption{Strict weak ordering of a partition class of IID nodes}\label{alg:order_of_IID_partition}
    \begin{algorithmic}[1]
        \If{$|P.f(x)| < |Q.f(x)|$}{~\textbf{return} \code{true}} \EndIf
        \If{$|P.f(x)| > |Q.f(x)|$}{~\textbf{return} \code{false}} \EndIf
        \State $W := [ 2^0, \ldots , 2^{10} ]$
        \State $p := \arg \underset{i}{\min}\{ | P.nbytes - W[i] | ~|~ i = 0, \ldots, |W|-1 \}$
        \State $q := \arg \underset{i}{\min}\{ | Q.nbytes - W[i] | ~|~ i = 0, \ldots, |W|-1 \}$
        \State $m := \arg \underset{i}{\min}\{ | max\_bytes/2 - W[i] | ~|~ i = 0, \ldots, |W|-1 \}$
        \If{$|W[m] - W[p]| < |W[m] - W[q]|$}{~\textbf{return} \code{true}} \EndIf
        \If{$|W[m] - W[p]| > |W[m] - W[q]|$}{~\textbf{return} \code{false}} \EndIf
        \If{$P.nbytes < Q.nbytes$}{~\textbf{return} \code{true}} \EndIf
        \If{$P.nbytes > Q.nbytes$}{~\textbf{return} \code{false}} \EndIf
        \State \textbf{return} $dP < dQ$
    \end{algorithmic}
\end{algorithm}

Once we have the representative pivot, say $P$, selected, then we may focus on
searching for a node $M$ (as explained above). First we should realize the following
facts which basically justify the approach we take.
\begin{itemize}
    \item There is no input $x$ to the \target{} such that the corresponding
        trace $T$ will be mapped to $P$ and the missing successor of $P$ will be
        created.
        \footnote{That is, of course, only true under the assumption the
        sensitivity analysis did not under-approximate the sensitive bits of
        $P$.}
        Therefore, if we want to cover $P$, then we have to escape from
        $\pth{P}$ at some node $\pth{P}[k]$, where $0 \leq k < dP - 1$.
    \item Although we want to escape from $\pth{P}$ at the index $k$, we still
        want to get back to the same instruction $P.id$. Although $P.id$ can be
        reached, in general, several completely different way, considering the
        information we have, it is reasonable to restrict our search for those
        paths which are similar to $\pth{P}$. There can be many of such similar
        paths. They may differ in the numbers of iterations in loops along the
        path. There may also be differences in what path is taken in each
        iteration of each loop. And observe that we indeed have valuable
        information about theses paths. Namely, for each IID pivot in the class
        $C$ we know the corresponding path. And for each of that path we also
        know the loops (entries, exits and bodies) along it.
\end{itemize}

So, our algorithm is as follows. We first need to go backwards along $\pth{P}$
(towards the root node) to find the index $k$ where to leave $\pth{P}$. Then, we
walk forward in the execution tree from the node $\pth{P}[k]$ along a path
similar to those in $C$. We will see, this forward walk is inspired in the Monte
Carlo walk used in games theory. Once we reach an open node $M$ with unexplored
success, which our similar path continues to, then we stop and $M$ is the node
we want an input generation analysis to be activated with.

Let us no focus on the computation of the index $k$. It is based on the following
observation in our evaluations:
\begin{itemize}
    \item The value of the branching function of an IID pivot typically depends
        on those loops (their iteration counts and interleaving of its path)
        which are close to the pivot. Higher the distance from the pivot, lower
        the chance of affecting the branching function. 
\end{itemize}
So, we thus choose $k$ as an index of a loop entry along $\pth{P}$. And we
should prefer those loop entries which are close to $P$. We collect all loop
entires along $\pth{P}$ using the \ralg{alg:detect_loops}. Formally, if $loops$
is the output from the algorithm, then we build the sequence $E = [ L.E ~|~E \in
loops ]$ of loop entries. Then we sort nodes in $E$ by their depth in the tree
in the decreasing order (because we want nodes closer to $P$ earlier in the
sequence). Next, we choose an index $i$ into $E$ randomly using the same
probabilities $p_i$ assigned to indices, which we used for selection of $P$ from
$C$. Lastly, our index $k$ is then $dE[i]$. Note though, that if $\pth{P}[k]$ is
closed, then we keep incrementing $k$ until $\pth{P}[k]$ is not closed. In case
the root node of the tree is closed, we cannot select any node in the tree, and
so we return \code{null}.

It remains to discuss how we describe a path similar to those in $C$ (used in
the Monte Carlo walk). The path is represented by two maps $\mathcal{F}$ and
$\mathcal{G}$, both from unique IDs of \code{Boolean} instructions, i.e.,
$\mathcal{F} = \mathcal{G} = \{ N.id.id ~|~ N \in \pth{N'} \wedge N' \in C \}$.

Given a node $N$, the value $\mathcal{F}(N.id.id)$ is the probability (in
$[0,1]$) with which we should move to $N.succ[\code{false}]$ (the probability to
move to $N.succ[\code{true}]$ is $1 - \mathcal{F}(N.id.id)$). The effectivity of
the Monte Carlo walk thus highly depends on these values. We compute them from
pivots in $C$. But not from all. We consider only this sequence $C' = [ N ~|~ N
\in C \wedge N.nbytes = P.nbytes ]$. And we sort it by the absolute value of the
branching function, i.e., by $|f(x)|$. This restriction to pivots to those
operating on inputs of the same size is important, because paths to them in the
execution tree tend to be highly similar, which leads to more accurate values in
$\mathcal{F}$ (than if all pivots were considered). We ordered $C'$, because
$|f(x)|$ also affects the final probability. Namely, for each $id.id$ and $0
\leq i < |C'|$, let $n_f(id.id, C'[i])$ be the count of nodes along
$\pth{C'}[i]$ with this $id.id$ and the the path continues from them to the
\code{false} successor. Similarly, let $n_t(id.id, C'[i])$ be the count to
\code{true} successors. And finally, let $F(id.id, C'[i]) = n_f(id.id, C'[i]) /
(n_f(id.id, C'[i]) + n_t(id.id, C'[i]))$. We then set $\mathcal{F}(id.id)$ to
the average of all these values
$$\{F(id.id, C'[i]) + t(F(id.id, C'[0]) - F(id.id, C'[i])) ~|~ 0 \leq i < |C'| \}$$
where $t = -|C'[i].f(x)| / (|C'[0].f(x)| - |C'[i].f(x)|)$, only with the nonzero
denominator, of course. If this set is empty, ten we put there $F(id.id, C'[0])$
and if $id.id$ was further detected to lie inside a loop body, then we also
include the value 0.5. Expression inside the set, together with the expression for $t$,
represent the solution on the following system
$$
    \begin{pmatrix} 0 \\ p \end{pmatrix}
    =
    \begin{pmatrix} |C'[i].f(x)| \\ F(id.id, C'[i]) \end{pmatrix}
    +
    t \begin{pmatrix} |C'[0].f(x)| - |C'[i].f(x)| \\ F(id.id, C'[0]) - F(id.id, C'[i]) \end{pmatrix}
$$
where the only unknown $p$ represents the element of the set. Due to ordering of
$C'$, the pivot $C'[0]$ is the one with the smallest $|f(x)|$ and it is thus the
closest node to the coverage of the corresponding \code{Boolean} instruction.
The processed pivot $C'[i]$ can be worse, so we interpolate the count $F(id.id,
C'[i])$ along the line in the right-hand side of the system. The purpose for the
addition of 0.5 to the set in the case $id.id$ being inside loop body is that we
actually want to add variability inside loop bodies in order to explore diverse
paths in loop iterations.

The range $Rng(\mathcal{G})$ consists of three random generators:
\begin{itemize}
    \item $G_{uni}$: Generates numbers in $[0,1]$ using the uniform
        distribution.
    \item $G_{1,0}$: This generator is initialized with the probability
        $\mathcal{F}(id.id)$ and the count $K = \sum_{i=0}^{|C'|-1}{n_f(id.id,
        C'[i]) + n_t(id.id, C'[i])}$. The generator then repeats the sequence of
        $K \mathcal{F}(id.id)$ numbers 1 and then $K - K \mathcal{F}(id.id)$
        numbers 0, forever.
    \item $G_{0,1}$: Differs from $G_{f,t}$ such that it first generates the
        sequence of zeros.
\end{itemize} 
The purpose of the last two generators is that the top one has difficulties to
generate the ``extreme'' sequences of the other two in reasonable time. However,
the extreme sequences are in fact quite common when speaking of paths in a loop.
The mapping of $Dom(\mathcal{G})$ to these generators is straightforward. If
$id.id$ represent a \code{Boolean} instruction inside a loop body, then we
choose randomly between all three generators. Othervise, we set
$\mathcal{G}(id.id)$ to $G_{uni}$.

With the maps $\mathcal{F}$ and $\mathcal{G}$ in hand the Monte Carlo walk
proceeds as follows. Let $N$ be the current node in the walk. Then the
\code{Boolean} value $b = \mathcal{F}(N.id.id) < \mathcal{G}(N.id.id)()$
\footnote{If $N.id.id \not\in Dom(\mathcal{F})$, then $\mathcal{F}(N.id.id)$
means $0.5$ and $\mathcal{G}(N.id.id)$ means $G_{uni}$.}
represents the direction in which we want to continue in the walk. However, we
can move in that direction only if $N.succ[b]$ is a valid non-closed node. If we
cannot move in the direction $b$ and $N$ is open, then we stop, because $N$ is
the node $M$ we have been searching for. Otherwise, we move into a non-closed
successor node (there has to be one), and we process it the same way.

\subsection{Recovery from early termination}\label{sec:recovery_from_termination}

Whenever the (typed) minimization fails to invert the result of evaluation of
the \code{Boolean} instruction corresponding to the processed node $N$, then
record the node together with the current value $n$ of the fuzzing loop's
iteration counter.

Later, when the fuzzing loop is being terminated with the reason that no node in
the tree can be selected for an input generation analysis, then we try to make
the recorded nodes available for the analysis.

So, let us consider our recorded node $N$. If $N.id$ was covered since it was
recorder, then we do nothing. We also do nothing, if $n \geq N.fn$. Otherwise,
we try to make the node available, because the (typed) minimization analysis may
succeed now. That is because the input $N.x$ has changed since the minimization
run with the node. Although it failed for the previous input it may succeed for
the current one.

To make $N$ available, we set all fields $sa$, $ba$, and $ma$ to \code{false}.
We also clear its closed state (if marked as such) and lastly we try to insert
it into the sets of the primary targets (see \rsec{sec:primary_targets}).

\section{Optimizer}\label{sec:optimizer}

After the termination of the fuzzing loop we look into generated tests. For each
test, for which the termination result was \code{BOUNDARY\_CONDITION\_VIOLATION}
we run the \target{} for the corresponding input again, but this time with all
limits highly extended. If the accepted trace improves the coverage of
\code{Boolean} instruction, then the accepted input is included to the final
test suite, if it actually differs from the input passed to the \target{}. 

\end{document}